\documentclass[%
 reprint,
superscriptaddress,
 amsmath,amssymb,
 aps,
 prd,
]{revtex4-2}

\usepackage{graphicx}
\usepackage{xcolor}
\usepackage{ulem}
\usepackage{dcolumn}
\usepackage{bm}
\usepackage{hyperref}
\usepackage[caption=false]{subfig}
\usepackage{siunitx}
\usepackage[capitalise]{cleveref}

\usepackage{lineno}

\begin{document}


\title{Probing gluon helicity with heavy flavor at the EIC}

\author{Daniele Paolo Anderle}
\email{dpa@m.scnu.edu.cn}
\affiliation{Guangdong Provincial Key Laboratory of Nuclear Science,\\ Institute of Quantum Matter, South China Normal University, Guangzhou 510006, China}
\affiliation{Guangdong-Hong Kong Joint Laboratory of Quantum Matter, Southern Nuclear Science Computing Center, South China Normal University, Guangzhou 510006, China}
\affiliation{Department of Physics and Astronomy, University of California, Los Angeles, California 90095, USA}

\author{Xin Dong}
\affiliation{Lawrence Berkeley National Laboratory, Berkeley, CA 94720, USA}

\author{Felix Hekhorn}
\email{felix.hekhorn@unimi.it}
\affiliation{Tif Lab, Dipartimento di Fisica, Università di Milano and\\ INFN, Sezione di Milano, Via Celoria 16, I-20133 Milano, Italy}

\author{Matthew Kelsey}
\affiliation{Wayne State University, Detroit, MI 48202, USA}
\affiliation{Lawrence Berkeley National Laboratory, Berkeley, CA 94720, USA}

\author{Sooraj Radhakrishnan}
\affiliation{Kent State University, Kent, OH 44242, USA}
\affiliation{Lawrence Berkeley National Laboratory, Berkeley, CA 94720, USA}

\author{Ernst Sichtermann}
\affiliation{Lawrence Berkeley National Laboratory, Berkeley, CA 94720, USA}

\author{Lei Xia}
\affiliation{University of Science and Technology of China, Hefei, Anhui Province 230026, China}

\author{Hongxi Xing}
\affiliation{Guangdong Provincial Key Laboratory of Nuclear Science,\\ Institute of Quantum Matter, South China Normal University, Guangzhou 510006, China}

\author{Feng Yuan}
\affiliation{Lawrence Berkeley National Laboratory, Berkeley, CA 94720, USA}

\author{Yuxiang Zhao}
\email{yxzhao@impcas.ac.cn}
\affiliation{Institute of Modern Physics, Chinese Academy of Sciences, Lanzhou, Gansu Province 730000, China}
\affiliation{University of Chinese Academy of Sciences, Beijing 100049, China}
\affiliation{Key Laboratory of Quark and Lepton Physics (MOE) and Institute of Particle Physics, Central China Normal University, Wuhan 430079, China}

\date{\today}

\begin{abstract}
We propose a new measurement of the heavy flavor hadron double spin asymmetry in deep-inelastic scattering at a future Electron-Ion Collider (EIC) to constrain
the polarized gluon distribution function inside the proton. 
Statistical projection on $D^0$ meson double spin asymmetry is calculated with an EIC central detector using an all-silicon tracker and vertexing subsystem.
A first impact study was done by interpreting pseudo-data at 
next-to-leading order in QCD. 
The sensitivity of the experimental observable in constraining gluon helicity distribution in a wide range of parton momentum fraction $x$ has been investigated considering different beam energy configurations.
This measurement complements the inclusive spin-dependent structure function measurement and provides an opportunity to constrain the gluon helicity distribution in the moderate $x$ region.

\end{abstract}

\maketitle

\section{Introduction}
The spin structure of the nucleon has been of fundamental interest in modern hadronic physics ever since the EMC spin puzzle~\cite{Ashman:1987hv}. The current understanding of the structure of the nucleon
spin is that it consists of the contributions from quark and gluon 
helicity distributions, as well as their orbital angular momenta~\cite{Ji:1996ek,*Wakamatsu:2010qj,*Ji:2012sj,Jaffe:1989jz}.
For a longitudinally polarized nucleon, its spin can be decomposed into~\cite{Jaffe:1989jz}
\begin{equation}
    \frac 1 2 = \frac{1}{2} \int\limits_0^1 \Delta \Sigma (x)\,dx + \int\limits_0^1 \Delta g(x)\,dx + L_q + L_g,
    \label{eq:Jaffe}
\end{equation}
where $\Delta \Sigma(x)$ and $\Delta g(x)$ denote the non-perturbative
(longitudinally-)polarized Parton Distribution Functions (PDFs) for the quark singlet and gluons, $L_q$ and $L_g$ are
the orbital angular momenta of quarks and gluons, and $x$ is the momentum fraction carried by the quarks or gluons.
The dependence of the PDFs on the factorization scale is left implicit here and for the most part in the rest of the paper.

After more than 40 years of both experimental and theoretical efforts, the precision of the unpolarized PDFs is reaching 
higher and higher accuracy~\cite{Hou:2019efy,*Ball:2017nwa,*Bailey:2020ooq,*Ball:2021leu}, but the polarized PDFs (pPDFs) are still not well 
constrained~\cite{Ethier:2020way}. In addition, compared to the polarized quark distributions the polarized gluon  
distribution $\Delta g$ is a less known quantity. 
This is because the gluon only participates via QCD evolution and higher-order corrections to fully inclusive Deep-Inelastic Scattering (DIS) that
most of the polarized experiments measure.
In the last decade or so, the polarized proton-proton experiments at
RHIC \cite{PHENIX:2004aoz,*PHENIX:2007kqm,*PHENIX:2008swq,*STAR:2009vxb,*PHENIX:2014gbf,*STAR:2006opb,*STAR:2007rjc,*STAR:2014wox} have provided a stronger constraint on the polarized gluon 
distribution due to the fact that $\Delta g$ enters the
differential cross section at the leading order in hard proton-proton scattering processes. 
Evidence for a positive $\Delta g$ in the region of $x>0.05$ has been found. Due to the limited kinematic coverage one
still cannot draw a decisive conclusion on the gluon spin
contribution to the proton spin~\cite{Ball:2018lag}.

The planned Electron-Ion Colliders (EIC)~\cite{Accardi:2012qut,Aschenauer:2017jsk,*AbdulKhalek:2021gbh,*Anderle:64701} offer unprecedented opportunities for spin physics, especially to constrain the $\Delta g$ contribution to the proton
spin via the scaling violations of the polarized structure functions~\cite{Accardi:2012qut}. 
Recently, there have been a series of studies to demonstrate the EIC impact on the pPDFs through various processes
including semi-inclusive light hadron production~\cite{PhysRevD.102.094018,Zhou:2021llj} and jet production~\cite{Borsa:2020ulb,*Page:2019gbf}.
In this paper we provide a systematic study of charm quark production to constrain the polarized gluon distribution in a wide range of $x$ at the future EIC.
Compared to the inclusive DIS measurements, the polarized charm structure function provides direct access to the $\Delta g$ from leading order. This will complement the fully inclusive DIS measurements in several important ways, for example, offering a new ingredient on the $\Delta g$ determination in addition to the inclusive DIS and providing sensitivity in the moderate $x$ region.  

A similar measurement was proposed and performed by the COMPASS collaboration \cite{COMPASS:2012mpe}. However, the COMPASS measurement 
yielded only one data point and was interpreted based on the approximation of photo-production~\cite{Bojak:2000eu}. 

Our study is based on two important developments in the recent years.
First, the next-to-leading order perturbative QCD formalism for
heavy flavor production in polarized DIS has been derived~\cite{Hekhorn:2018ywm,*Hekhorn:2019nlf,*Hekhorn:2021cjd}. This will help to achieve high
precision from theory side in constraining $\Delta g$ from the experiments.
Second, an all-silicon tracker conceptual design has been applied
and demonstrated in various EIC simulations~\cite{Arrington:2021yeb,Kelsey:2021gpk}. It also plays an essential role in our analysis of this paper, since it enables high precision measurement of heavy flavor hadrons through their hadronic decay channels.

\section{theoretical calculation}
\label{section:theory}
In the theory calculation, we focus on electro-production of inclusive open charm particles. The hadronization effects of the charm quark into the $D$ mesons, electro-weak corrections, intrinsic charm components, and target mass corrections are 
currently not considered.
The cross-sections for the unpolarized and polarized DIS processes are given in terms of three independent structure functions $F_{1,2}(x,Q^2)$ and $g_{1}(x,Q^2)$:
\begin{align}
    \frac{d^2\sigma}{d{x} dy} &= \frac{4\pi \alpha^2}{x y Q^2} \left[(1-y)F_2(x,Q^2) + y^2 x F_1(x,Q^2) \right],\\
    \frac{d^2 \Delta\sigma}{dx dy} &= \frac{4\pi \alpha^2}{x y Q^2} (2y-y^2) 2xg_1(x,Q^2)
\end{align}
from which we can define the charm double-spin asymmetry
\begin{align}
    A_{LL}^c 
    &=\frac{d\sigma^{++}-d\sigma^{+-}}{d\sigma^{++}+d\sigma^{+-}} = \frac{d\Delta \sigma}{2d\sigma} \\
    &\approx D(y)\frac{g_1^c(x,Q^2)}{F_1^c(x,Q^2)} \\
    &=\frac{y(2-y)}{y^2+2(1-y)}\frac{g_1^c(x,Q^2)}{F_1^c(x,Q^2)},
\end{align}
where the superscript $c$ refers to the charm component of the structure functions,
$d\sigma^{++}$ and $d\sigma^{+-}$ are the charm production cross-sections for electron and
proton beam spin orientation to be parallel and anti-parallel to each other, respectively,
and $D(y)$ is the depolarization factor of the virtual photon depending on the inelasticity $y$. The target mass as
well as the cross-section from longitudinal photon polarization are ignored
in the above equations.

In the context of collinear factorization, the structure functions can be computed as a convolution of (p)PDFs $(\Delta) f_j$ and perturbative coefficient functions $(\Delta) c_{k,j}$:
\begin{align}
    F_{[1,2]}^c(x,Q^2) &= \sum_{j=g,q,\bar q} \int\limits_x^{z_{max}} \frac{dz}{z} f_j\left(\frac{x}{z},\mu_F^2\right) c_{[1,2],j}(z,Q^2),\label{eq:F12}\\
    g_{1}^c(x,Q^2) &= \sum_{j=g,q,\bar q} \int\limits_x^{z_{max}}\frac{dz}{z}  \Delta f_j\left(\frac{x}{z},\mu_F^2\right) \Delta c_{1,j}(z,Q^2),\label{eq:g1}
\end{align}
where $z_{max} = Q^2/(4m^2+Q^2)$ is the kinematic boundary to create a charm quark pair in the final state with $m$ the charm quark mass. Note that the argument of the PDF is $x/z$ where $x$ is the Bjorken-$x$ and $z$ is the convolution variable.
The perturbative next-to-leading order (NLO) calculation of the partonic coefficient functions $(\Delta) c_{k,j}$ is known in the unpolarized case~\cite{Laenen1993162} for quite a while. Their polarized counterparts have become available only recently~\cite{Hekhorn:2018ywm,*Hekhorn:2019nlf,*Hekhorn:2021cjd} after the previous leading order (LO) computation~\cite{Watson:1981ce,*Gluck:1990in}. 

\begin{figure}[htbp]
\centering
\includegraphics[width=0.5\textwidth]{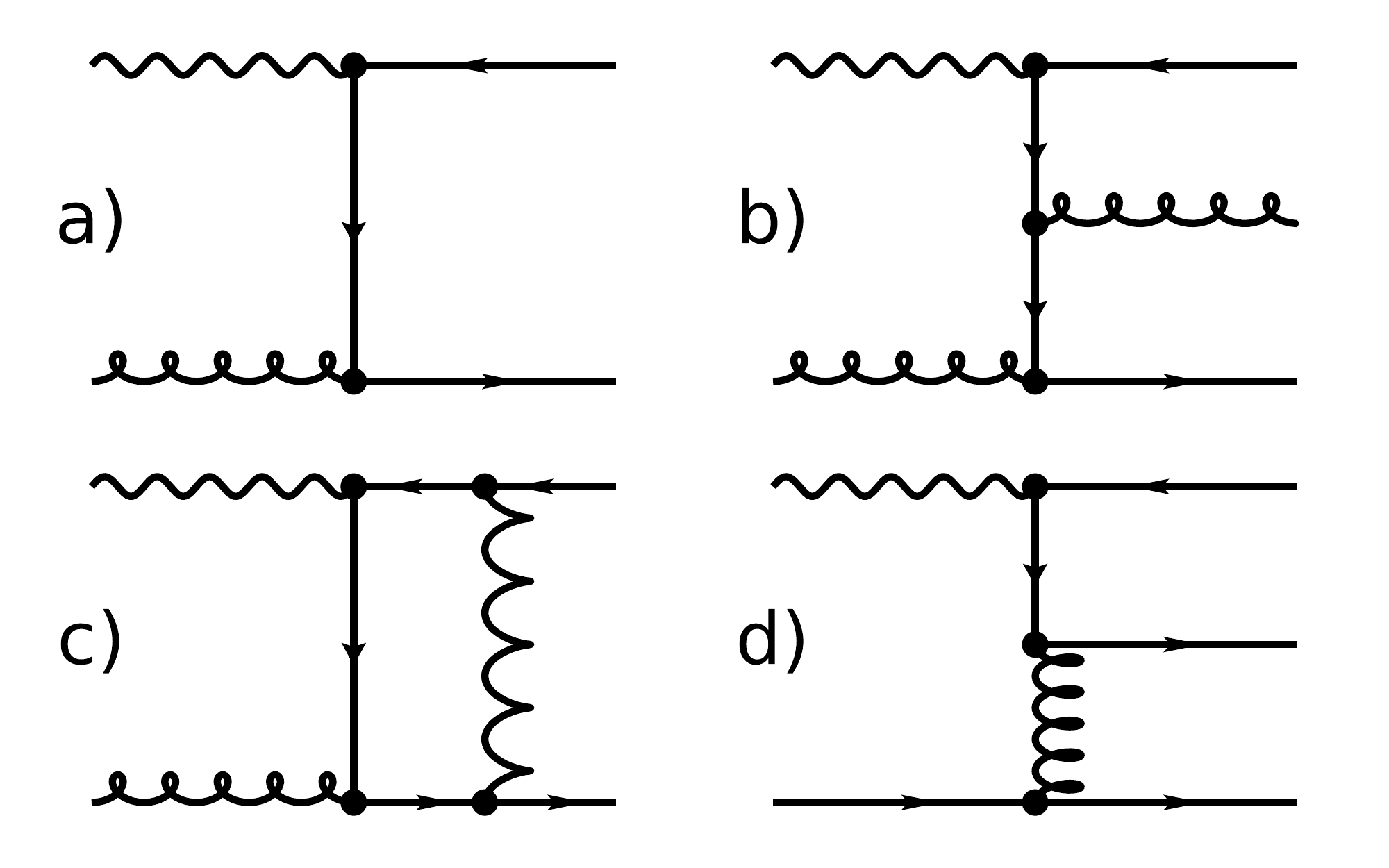}
\caption{Selected Feynman diagrams at LO and NLO for heavy flavor production in 
high energy $e+p$ collisions.}
\label{fig:Feyn}
\end{figure}

Heavy flavor production can constrain the gluon PDF since at LO the only contribution is photon-gluon-fusion (PGF) (see \cref{fig:Feyn}a) and in the case of unpolarized PDFs this is an established technique \cite{H1:2018flt}. At NLO three different types of contributions have to be considered: real gluon radiation (see \cref{fig:Feyn}b), virtual corrections to PGF (see \cref{fig:Feyn}c), and light quark initiated contributions (see \cref{fig:Feyn}d).

For the factorization and renormalization procedure we use the $\overline{\rm{MS}}_m$ scheme as described in \cite{Bojak:2000eu} and choose the respective scales to be $\mu_F^2 = \mu_R^2 = 4m^2 + Q^2$. We use a fixed flavor scheme with $n_l=3$ light flavors (uds) together with a pole mass prescription for the heavy charm quark. The actual value for the pole mass $m$ and the prescription for the running coupling $\alpha_s(\mu_R^2)$ is provided by the LHAPDF interface \cite{LHAPDF6}.
For the calculations of this paper, we use both the NNPDFpol1.1 \cite{Nocera:2014gqa} and  DSSV14~\cite{deFlorian:2014yva,*deFlorian:2019zkl} polarized PDF sets alongside with their unpolarized counterpart NNPDF23\_nlo~\cite{Ball:2012cx} and MSTW2008nlo~\cite{Martin:2009iq}, respectively.
\section{Projections for the experimental observable}
\begin{figure}[htbp]
\centering
\includegraphics[width=0.48\textwidth]{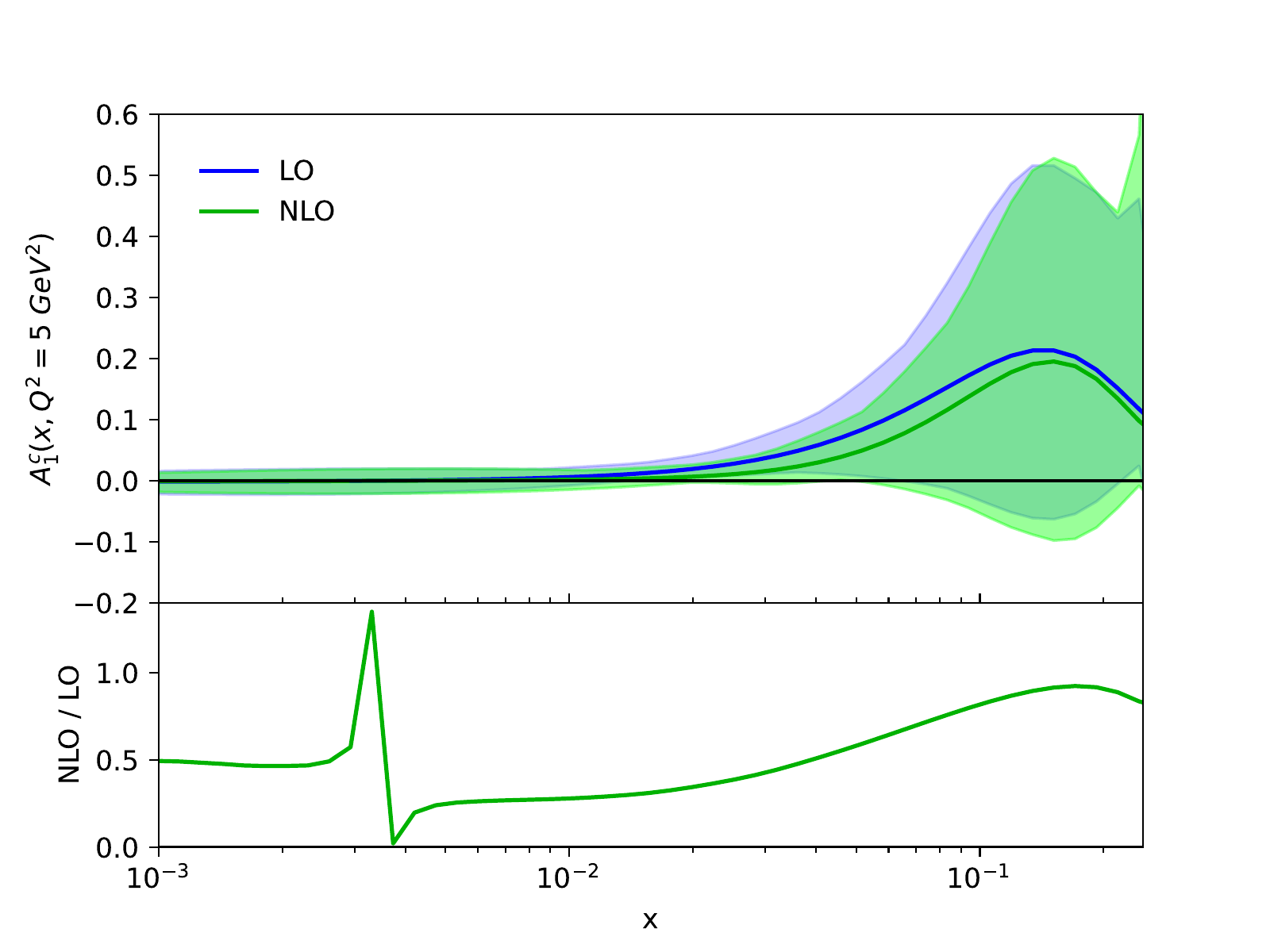}
\caption{Double-spin asymmetry $A_{1}^c$ as a function of $x$ for fixed virtuality $Q^2 = \SI{5}{\GeV^2}$ using the DSSV PDF set. In the upper panel both the LO predictions and the NLO predictions are shown with their respective PDF uncertainty. The lower panel shows the ratio between the calculations at NLO and LO accuracy.
Note that the predicted asymmetry $A_{1}^c$ at moderate $x$ region
is around 10-20\%.}
\label{fig:ALL}
\end{figure}

Experimentally, the double spin asymmetry in 
the $\vec{e}+\vec{p}\to e' + D^0 + X$ DIS process can be measured at an EIC as
\begin{align}
    A_{LL}^{\vec{e}+\vec{p} \to e' + D^0 + X} &=\frac{d\sigma^{++}-d\sigma^{+-}}{d\sigma^{++}+d\sigma^{+-}} \\
    &=\frac{1}{P_e P_p}\frac{N^{++}-N^{+-}}{N^{++}+N^{+-}}
\end{align}
where $N^{++}$ and $N^{+-}$ are the 
luminosity-normalized counts for electron and
proton beam spin orientation to be parallel and anti-parallel to each other, respectively, and $P_e$ ($P_p$) is the electron (proton) beam
polarization. 
The beam polarization is assumed to be 80\% for the electron beam and 70\% for the proton beam at the EIC~\cite{Accardi:2012qut}.
Therefore, one has
\begin{equation}
A_1^c\equiv\frac{g_1^c(x,Q^2)}{F_1^c(x,Q^2)} = \frac{1}{D(y)} \frac{1}{P_e P_p}\frac{N^{++}-N^{+-}}{N^{++}+N^{+-}},
\label{eq:A1}
\end{equation}
where $A_1^{c}$ can be calculated as discussed in the \cref{section:theory}.
To demonstrate the general size of the double-spin asymmetry $A_{1}^c$, we show a representative plot in \cref{fig:ALL}.

An all-silicon tracking detector design~\cite{Arrington:2021yeb} at an EIC enables the 
$D^0$ reconstruction with a very good signal-to-background ratio.
Moreover, the large acceptance and high luminosity available at an EIC allows the measurement to be done in 
a broad kinematic coverage in Bjorken-$x$ and $Q^2$. 

A simulation study has been performed to obtain uncertainty projections of the experimental observable~\cite{Arrington:2021yeb}.
The geometry of a silicon tracking system has been implemented in GEANT4 and studied within the full
Monte-Carlo framework for detector simulation. The full simulation yields the
detector response tables for momentum resolution, single track pointing resolution, 
tracking efficiency and primary vertex resolution. Afterward, the resolution tables were implemented in a fast smearing
simulation framework to allow for the generation of sufficient 
statistics to carry out detailed studies for physics projections. 
In our study, three beam energy configurations have been used for electron-proton 
collisions: $\SI{18}{\GeV} \times \SI{275}{\GeV}$, 
 $\SI{5}{\GeV}\times\SI{100}{\GeV}$  and $\SI{5}{\GeV}\times\SI{41}{\GeV}$.

The
data were generated by pythiaeRHIC (PYTHIA V6.4) and then fed into the fast smearing framework to accommodate detector
response within a \SI{3}{\tesla} magnetic field. We take advantage of the $K\pi$ two-body-decay to identify the $D^0$($\bar{D^0}$). 
Three decay
topological distributions, namely, $K\pi$ pair-DCA (distance of closest approach), $D^0$ Decay-Length$_{r\phi}$ in the transverse plane, and the 
cos$\theta_{r\phi}$ where $\theta$ is the angle
between the $D^0$ pointing direction with respect to the primary vertex and the 
momentum vector of the $K\pi$ pair, were investigated to obtain the data sample with a good signal-to-background ratio.
In addition to the $D^0$-decay topology cuts, the following kinematic cuts in the $e+p$ collision
including the squared momentum transfer of the electron $Q^2$, the inelasticity $y$, and
the invariant mass of the produced hadronic system $W$ were used in the analysis: $Q^2 > \SI{2}{\GeV^2}$, $0.05 < y < 0.8$, 
and $W^2 > \SI{4}{\GeV^2}$. The pion/kaon identification was assumed to be feasible up to the momentum limits \SIlist{10;6;50}{\GeV/c} in pseudo-rapidity regions
(-3,-1), (-1,1) and (1,3), respectively. 

\begin{figure*}[htbp]
\centering
\includegraphics[width=0.31\textwidth]{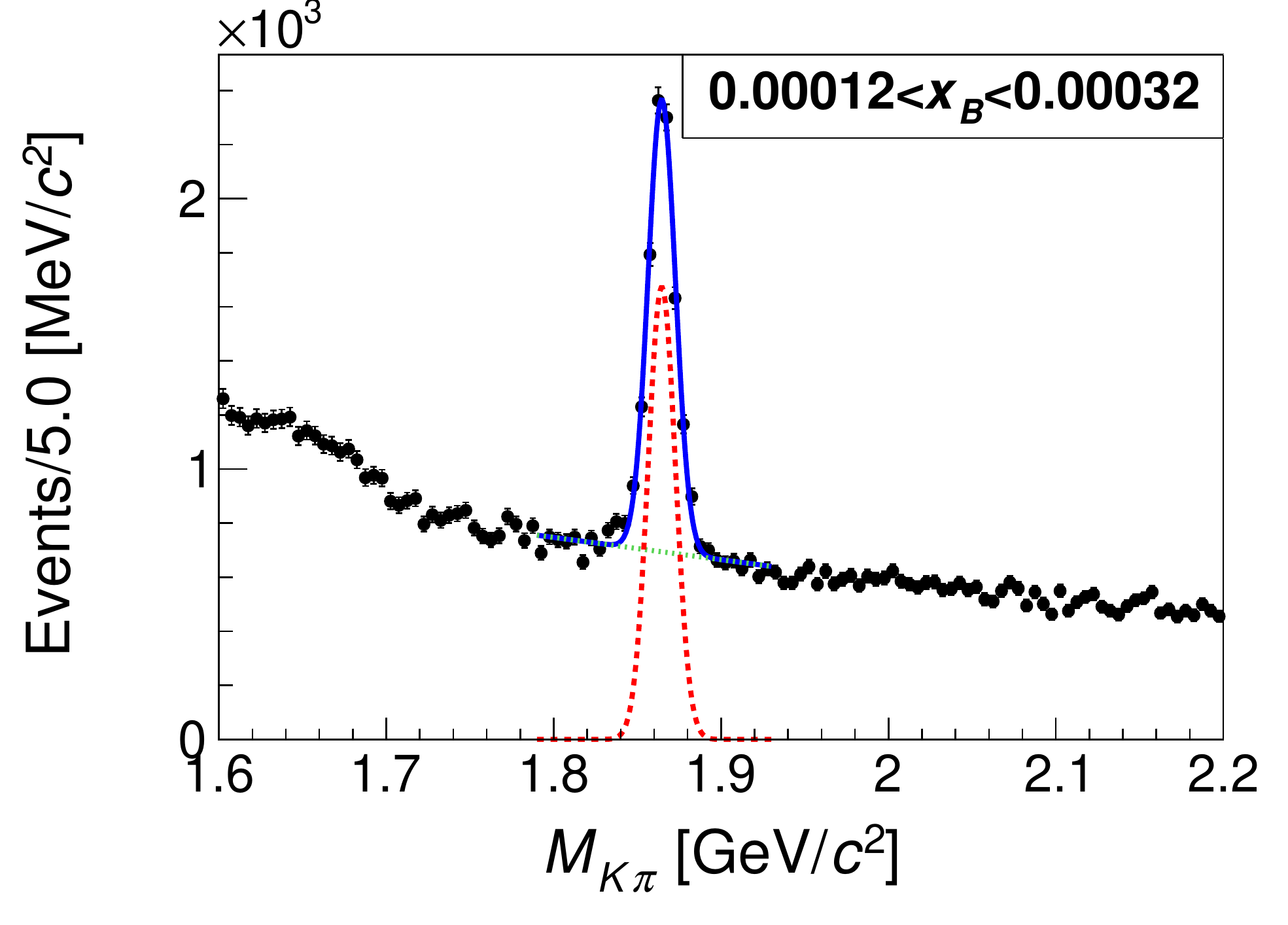}
\includegraphics[width=0.31\textwidth]{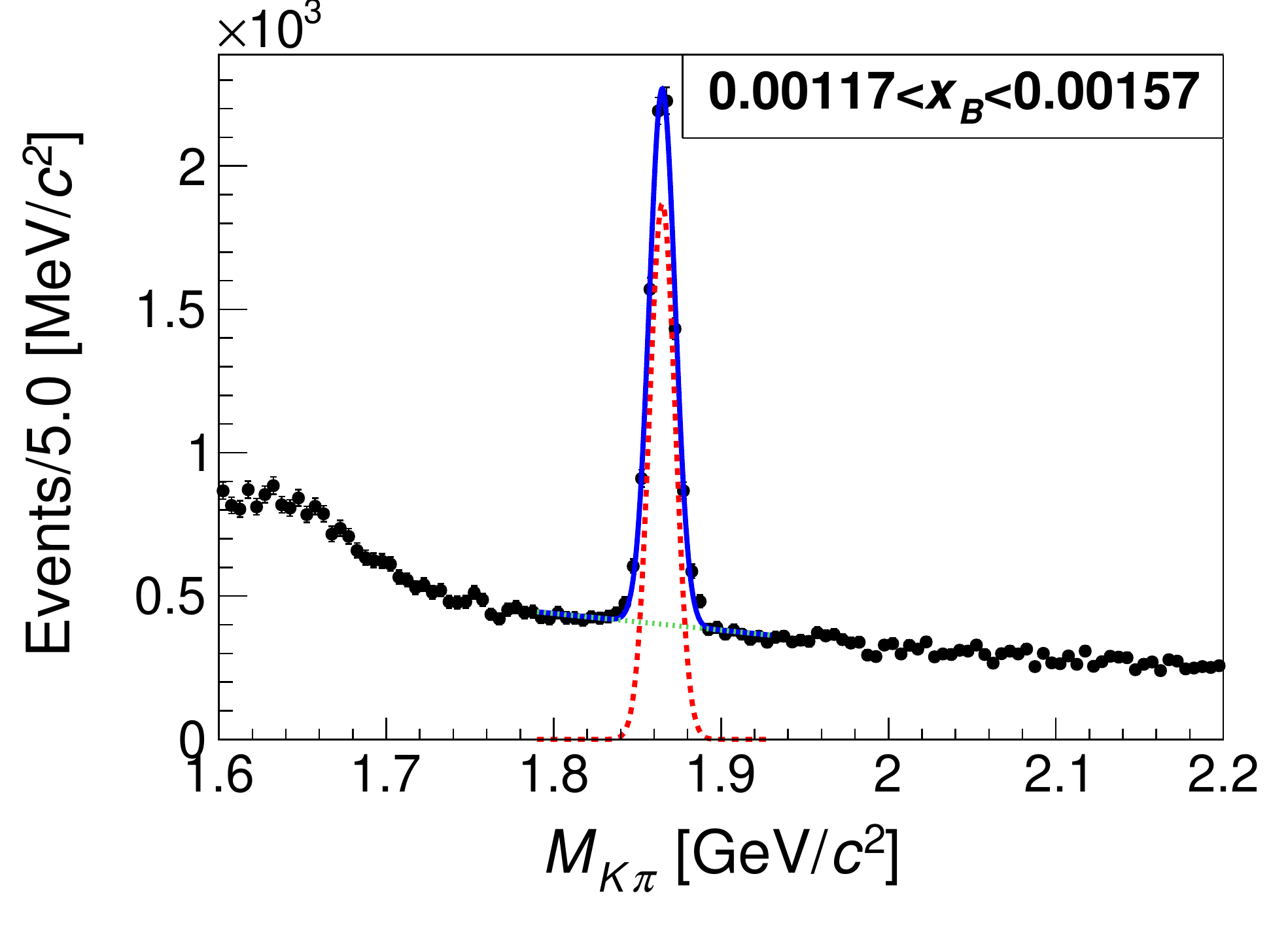}
\includegraphics[width=0.31\textwidth]{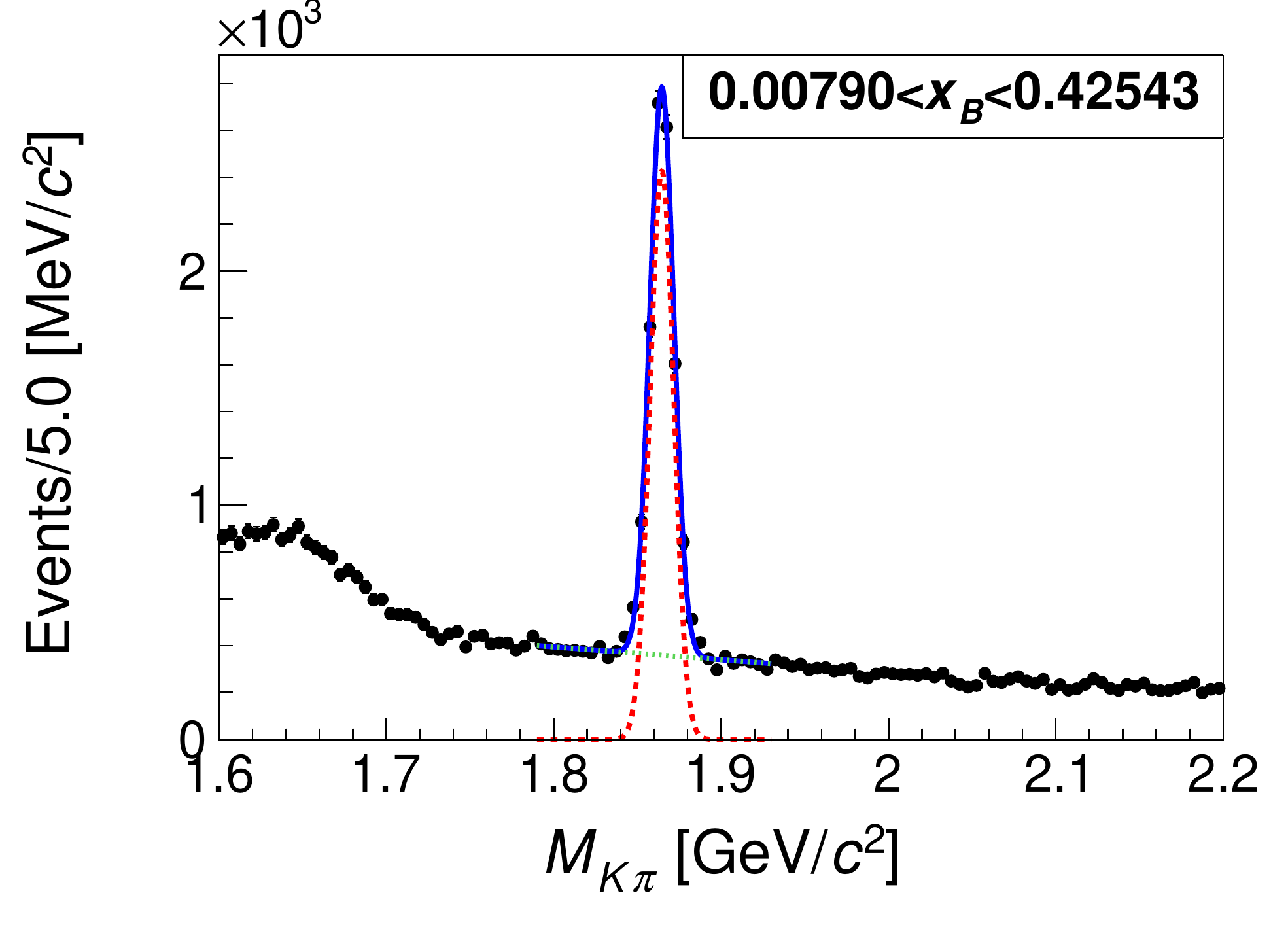}
\caption{(Color online) Fits to the $K\pi$ invariant-mass distributions in a few different Bjorken-$x$ bins for 18 GeV $\times$ 275 GeV $e$+$p$ collisions. The red and green dashed curves are the signal (Gaussian) and background (linear) fits, and the blue curve is the sum. 
}  
\label{fig_fit_18_275}
\end{figure*}

\begin{figure}[htbp]
\centering
\includegraphics[width=0.5\textwidth]{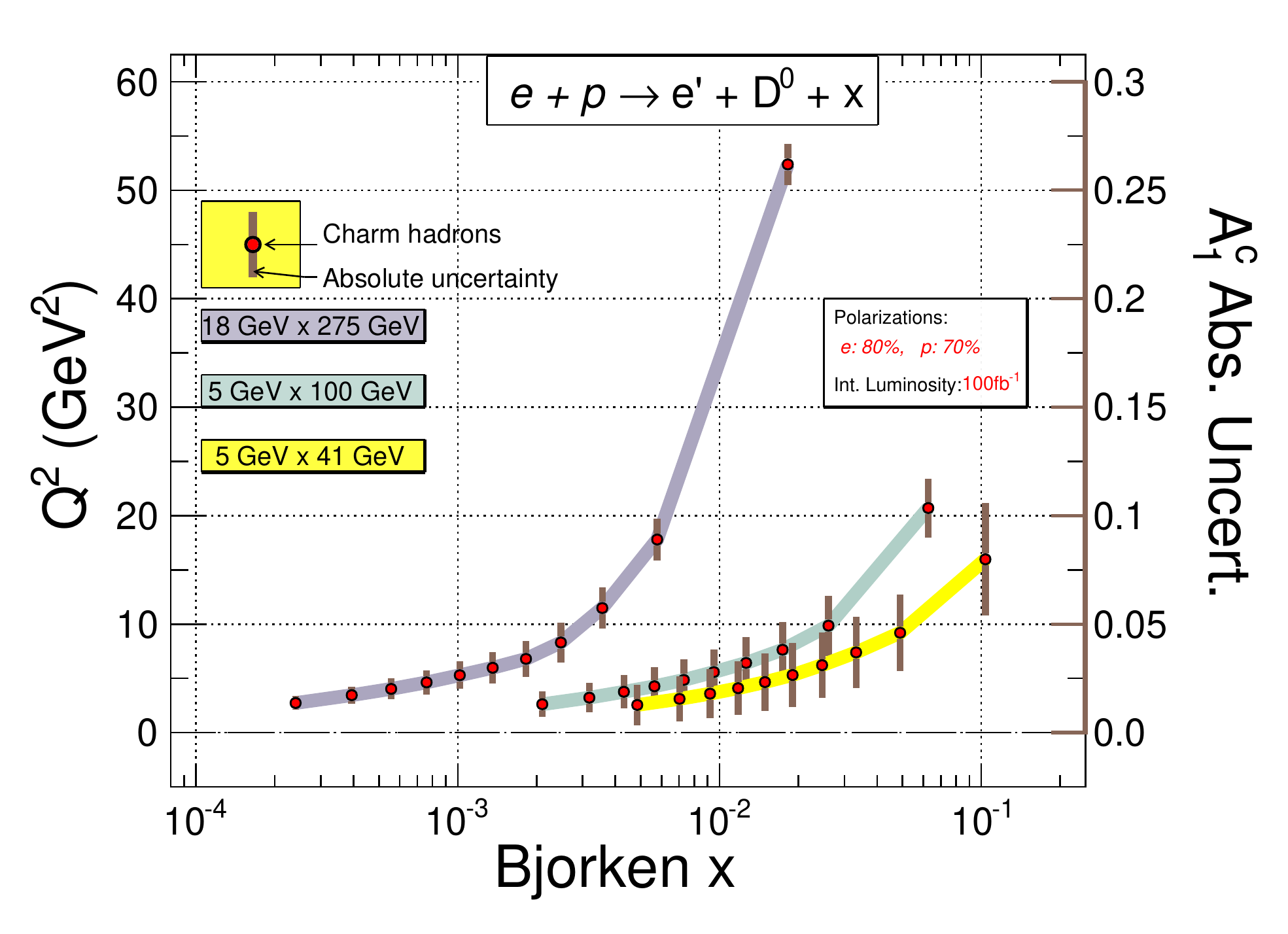}
\caption{(Color online) Projections of the double-spin asymmetry A$_{1}^{c}$ (formula \ref{eq:A1}) in the
$\vec{e}+\vec{p} \to e + D^0 + X$ process in bins of Bjorken-$x$ for different beam-energy configurations.
The integrated luminosity is \SI{100}{\per\femto\barn} for each configuration in this plot.
The electron (proton) beam polarization is assumed to be 80\% (70\%).
The position of each data point in the plot is defined by the
weighted center
of Bjorken-$x$ and Q$^2$ for each particular bin. The uncertainty indicated for each data point 
should be interpreted using the scale shown on the right-side vertical axis of the plot.}
\label{fig_ALL_projection}
\end{figure}

After all the selection requirements had been applied, the data were binned in Bjorken-$x$. In each bin the reconstructed
$K\pi$ invariant mass spectrum was fit with a Gaussian function for signal plus a linear background to extract the number of 
$D^0$ signal and background (as shown in \cref{fig_fit_18_275}). Hence, the uncertainty of $A_{1}^{c}$
can be calculated bin by bin, as shown in \cref{fig_ALL_projection}
for three different beam energy configurations. The weighted center for each data point is according to the Bjorken-$x$ and $Q^2$ axis, while the size
of the error is according to the scale on the right-side vertical axis. The integrated luminosity is corresponding to \SI{100}{\per\femto\barn} for
each beam energy configuration. The electron (proton) beam polarization
is assumed to 80\% (70\%).
The uncertainties become larger in the lower beam-energy configuration is due to the decrease of the production cross-section
for the charm quark.
For each beam energy configuration, the uncertainty
becomes larger in the higher $x$ region due to the smaller depolarization factor $D(y)$.

\begin{figure*}[ht]
\centering
\subfloat[With NNPDFpol1.1 PDFs\label{fig:impact_PDFs_NNPDF}]{\includegraphics[clip, trim=1cm 0cm 2cm 1cm,width=0.8\textwidth]{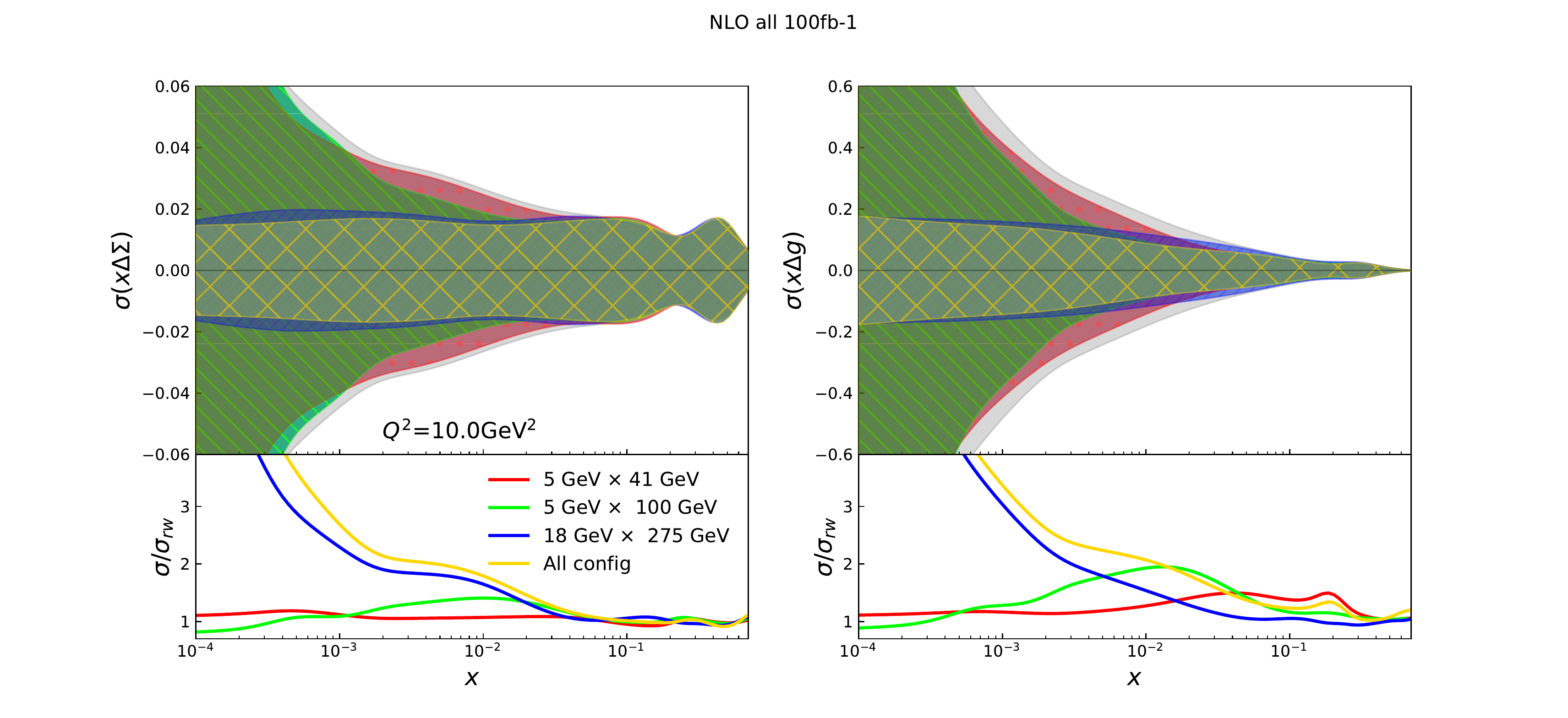}}
\\
\subfloat[With DSSV14 PDFs\label{fig:impact_PDFs_DSSV14}]{\includegraphics[clip, trim=1cm 0cm 2cm 1cm,width=0.8\textwidth]{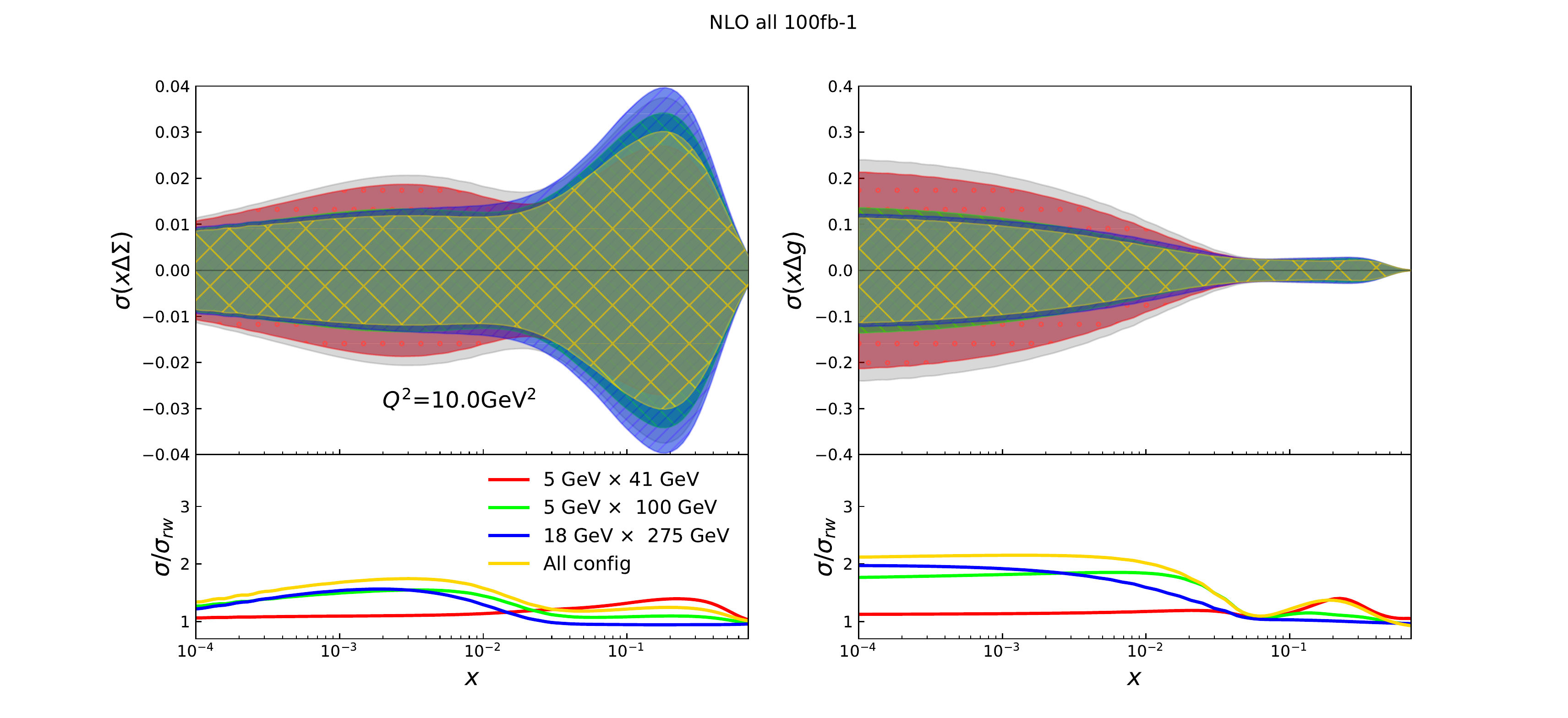}}
\\
\caption{\label{fig:impact_PDFs}(Color online) The quark singlet and gluon helicity distributions constrained by the $D^0$
double spin asymmetry pseudo-data in future EIC experiments at three different energies.
The top and bottom plots show the results by
using NNPDFpol1.1 and DSSV14 replicas. 
The top panels of each plot represent the absolute uncertainty of $x$ times the distribution:
the grey band shows the original uncertainty, the red (green, blue) band shows the updated uncertainty
by adding $\SI{5}{\GeV} \times \SI{41}{\GeV}$ ($\SI{5}{\GeV} \times \SI{100}{\GeV}$, $\SI{18}{\GeV} \times \SI{275}{\GeV}$) EIC pseudo-data.
The bottom panels of each plot show the ratio between the uncertainties before and after reweighting.
In addition, the resulted impact by including all three pseudo-data sets in 
the reweighting procedure is shown in yellow color. The PDFs are evaluated at $Q^2=\SI{10}{\GeV^2}$.
The integrated luminosity is \SI{100}{\per\femto\barn} for each beam 
energy configuration. }
\end{figure*}

\begin{figure*}[ht]
\centering
\subfloat[Nucleon spin contribution from quarks with $x>x_{\text{min}}$]{\includegraphics[clip, trim=0cm 0cm 2cm 1cm,width=0.48\textwidth]{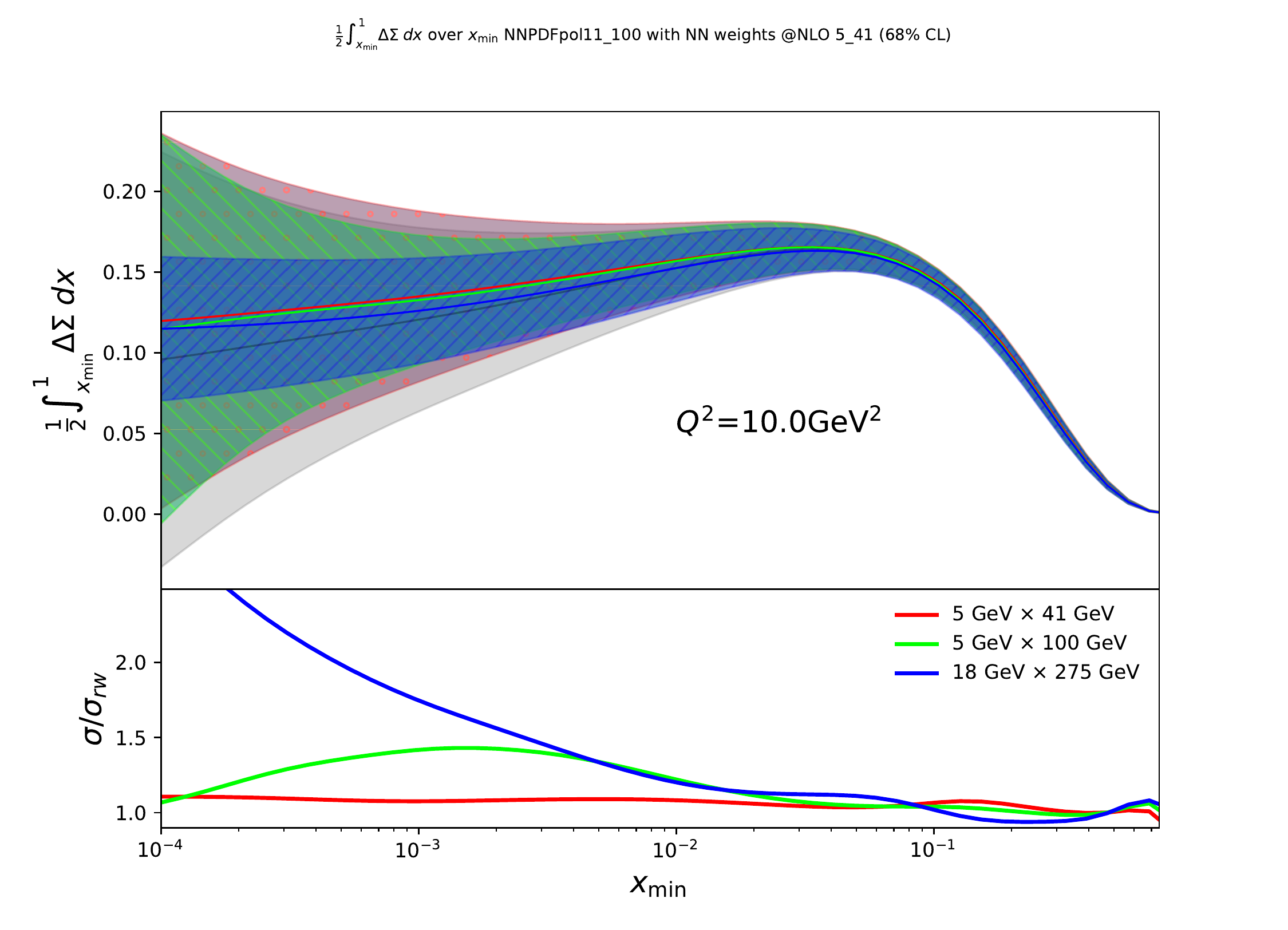}}
\subfloat[Nucleon spin contribution from gluons with $x>x_{\text{min}}$ ]{\includegraphics[clip, trim=0cm 0cm 2cm 1cm,width=0.48\textwidth]{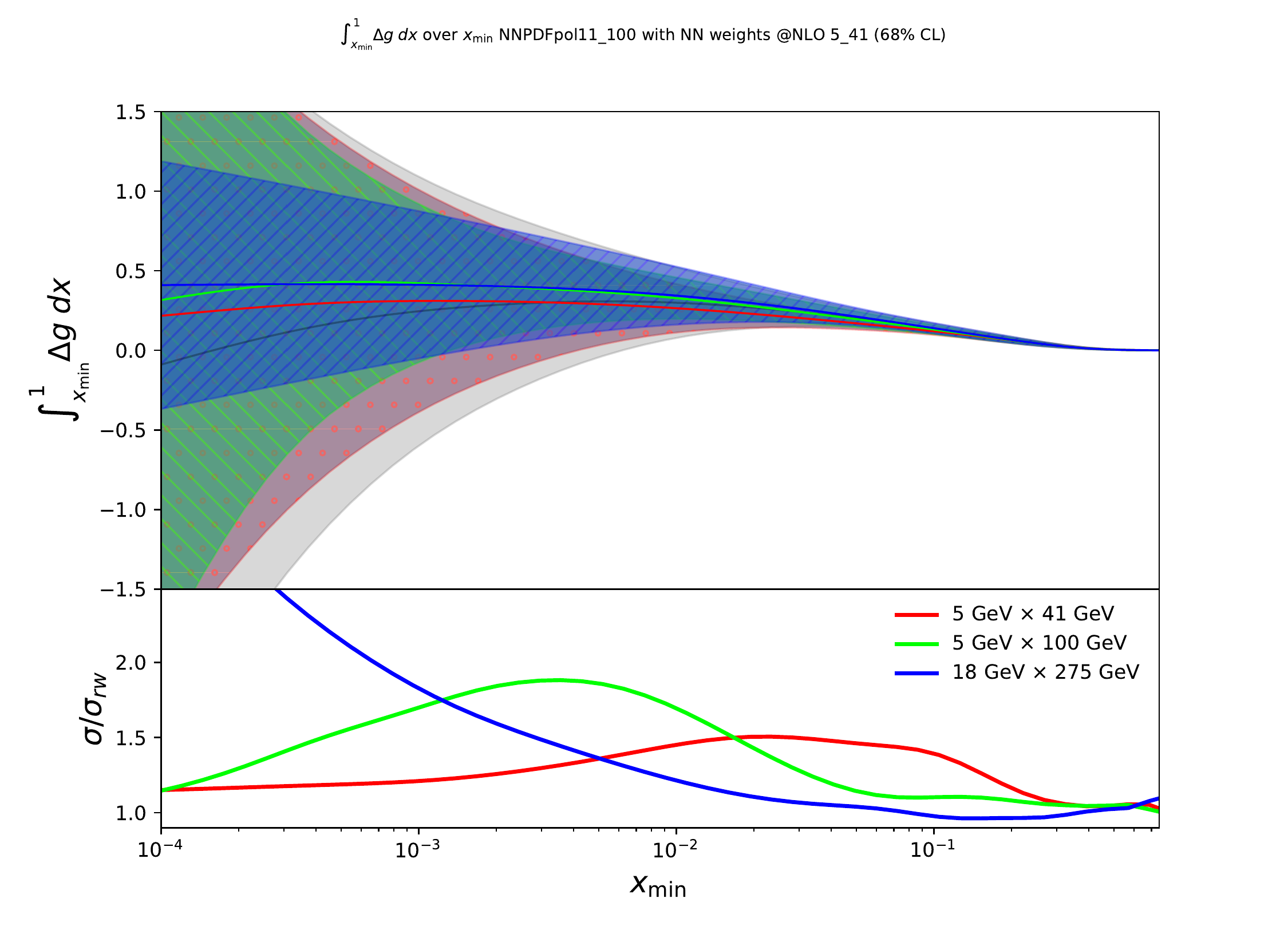}}
\\
\subfloat[Missing spin contribution to the nucleon from partons with $x>x_{\text{min}}$]{\includegraphics[clip, trim=0cm 0cm 2cm 1cm,width=0.49\textwidth]{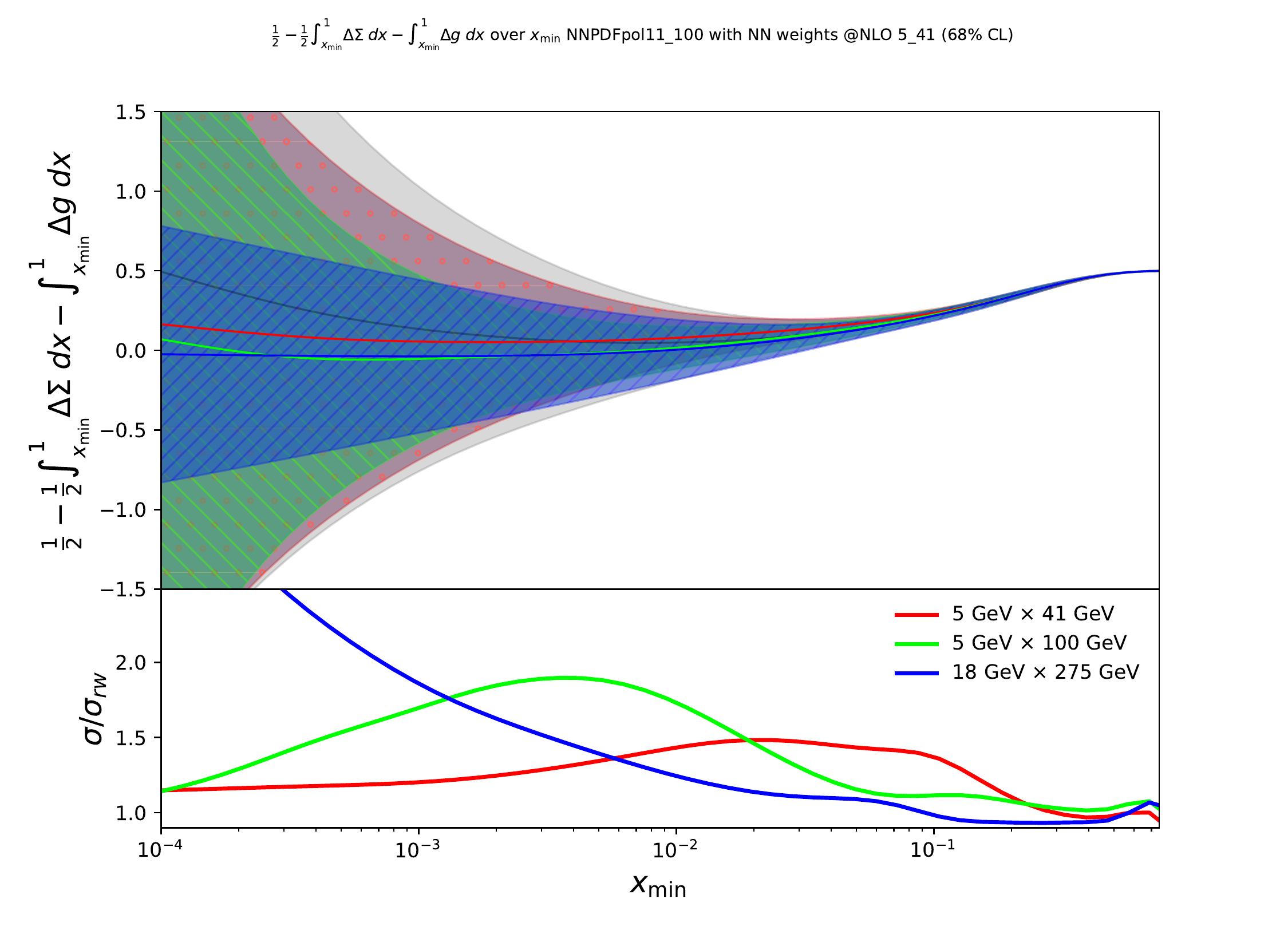}}
\\
\caption{(Color online) The top panel of each plot shows the result of the reweighting procedure to the integrals of singlet and gluon helicity NNPDFpol1.1 distributions as a function of the lower integration limit $x_{\text{min}}$. 
In addition, the contribution of the quark and gluon orbital angular momenta to
the proton spin is also shown. 
The integrated luminosity is \SI{100}{\per\femto\barn} for each beam 
energy configuration. The grey band shows the associated uncertainty according to the original NNPDFpol1.1 errorbands, the red (green, blue) band shows the updated uncertainty
by adding $\SI{5}{\GeV} \times \SI{41}{\GeV}$ ($\SI{5}{\GeV} \times \SI{100}{\GeV}$, $\SI{18}{\GeV} \times \SI{275}{\GeV}$) EIC pseudo-data.
The lower panel of each plot shows the ratio between the uncertainties before and after reweighting. The PDFs are evaluated at $Q^2=\SI{10}{\GeV^2}$. }
\label{fig:xmin_NNPDF}
\end{figure*}

\begin{figure*}[ht]
\centering
\subfloat[Nucleon spin contribution from quarks with $x>x_{\text{min}}$]{\includegraphics[clip, trim=0cm 0cm 2cm 1cm,width=0.48\textwidth]{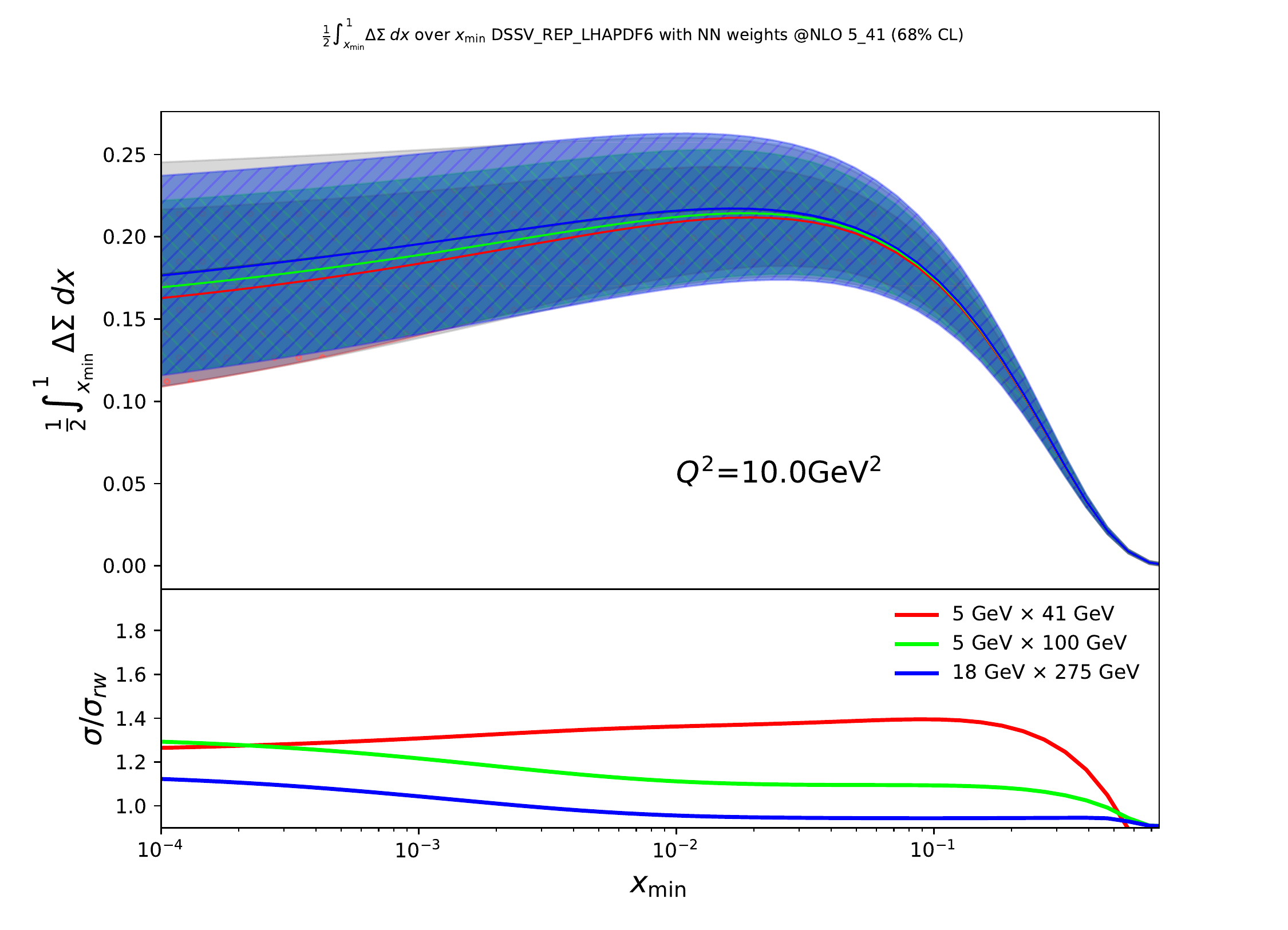}}
\subfloat[Nucleon spin contribution from gluons with $x>x_{\text{min}}$ ]{\includegraphics[clip, trim=0cm 0cm 2cm 1cm,width=0.48\textwidth]{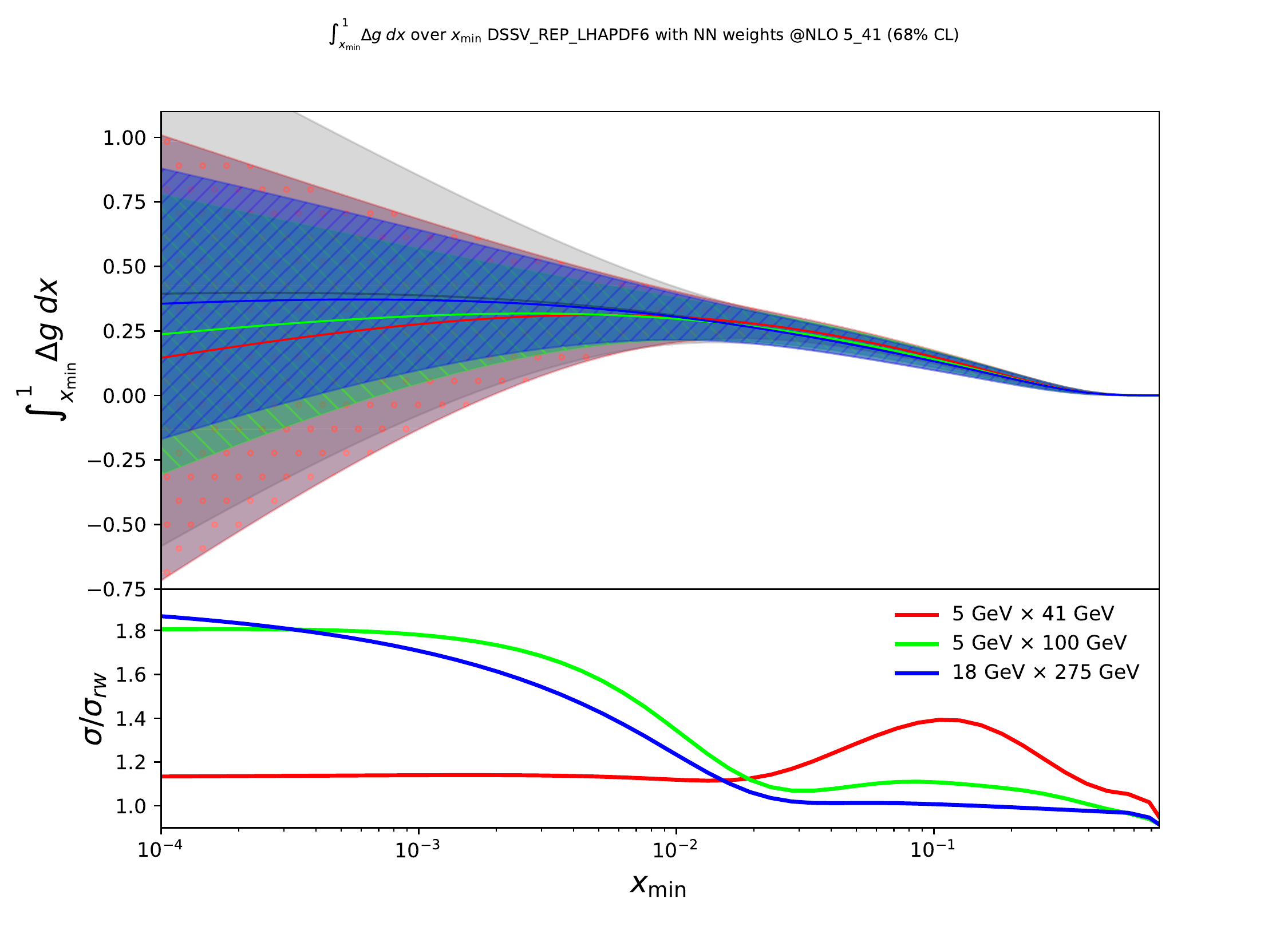}}
\\
\subfloat[Missing spin contribution to the nucleon from partons with $x>x_{\text{min}}$]{\includegraphics[clip, trim=0cm 0cm 2cm 1cm,width=0.49\textwidth]{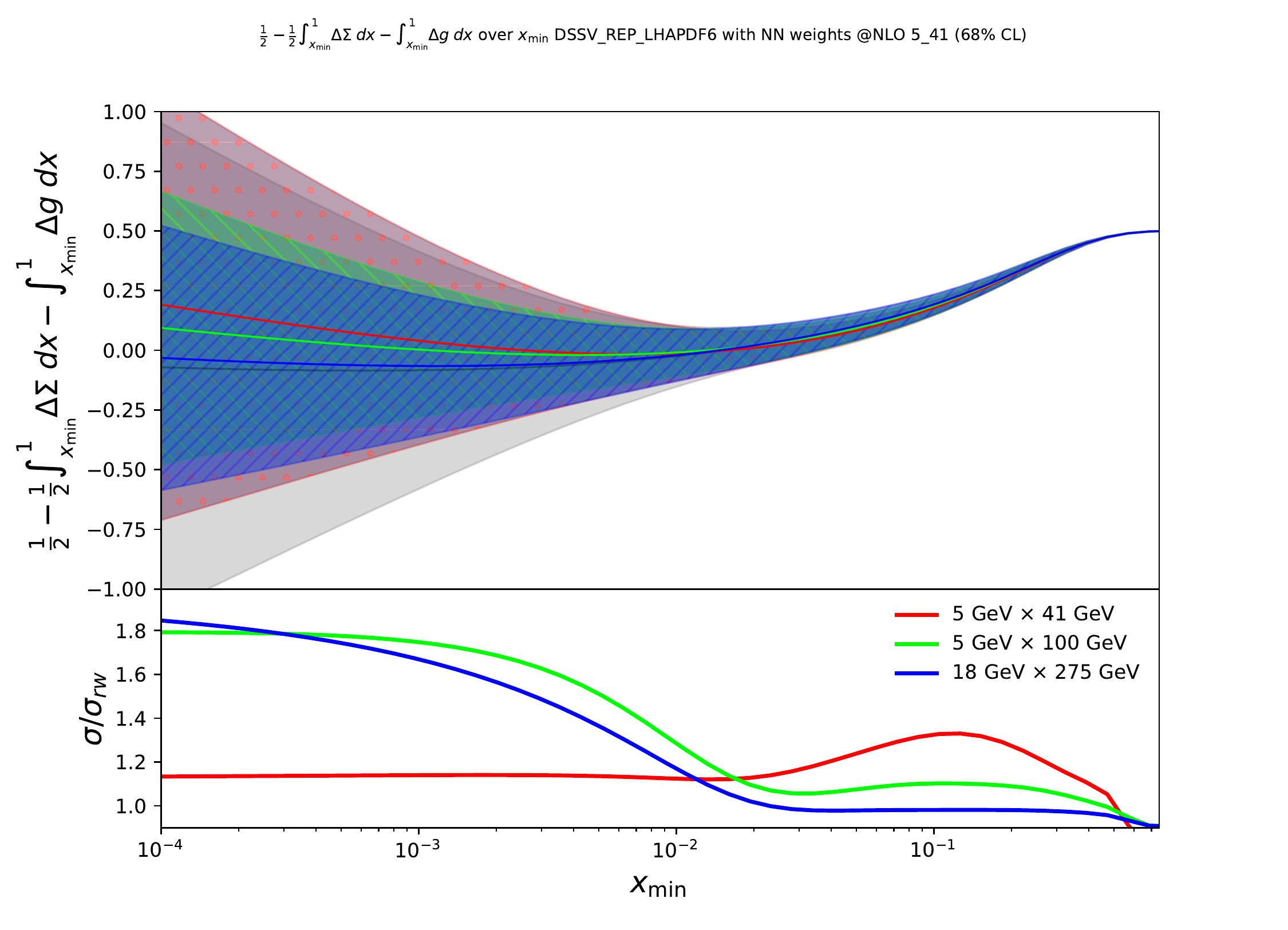}}
\\
\caption{(Color online) Same as \cref{fig:xmin_NNPDF} for the DSSV14 distributions. }
\label{fig:xmin_DSSV}
\end{figure*}

\begin{figure}[htbp]
\centering
\subfloat[For $\SI{5}{\GeV} \times \SI{41}{\GeV}$ energy configuration]{\includegraphics[clip, trim=0.5cm 0cm 2cm 1cm,width=0.45\textwidth]{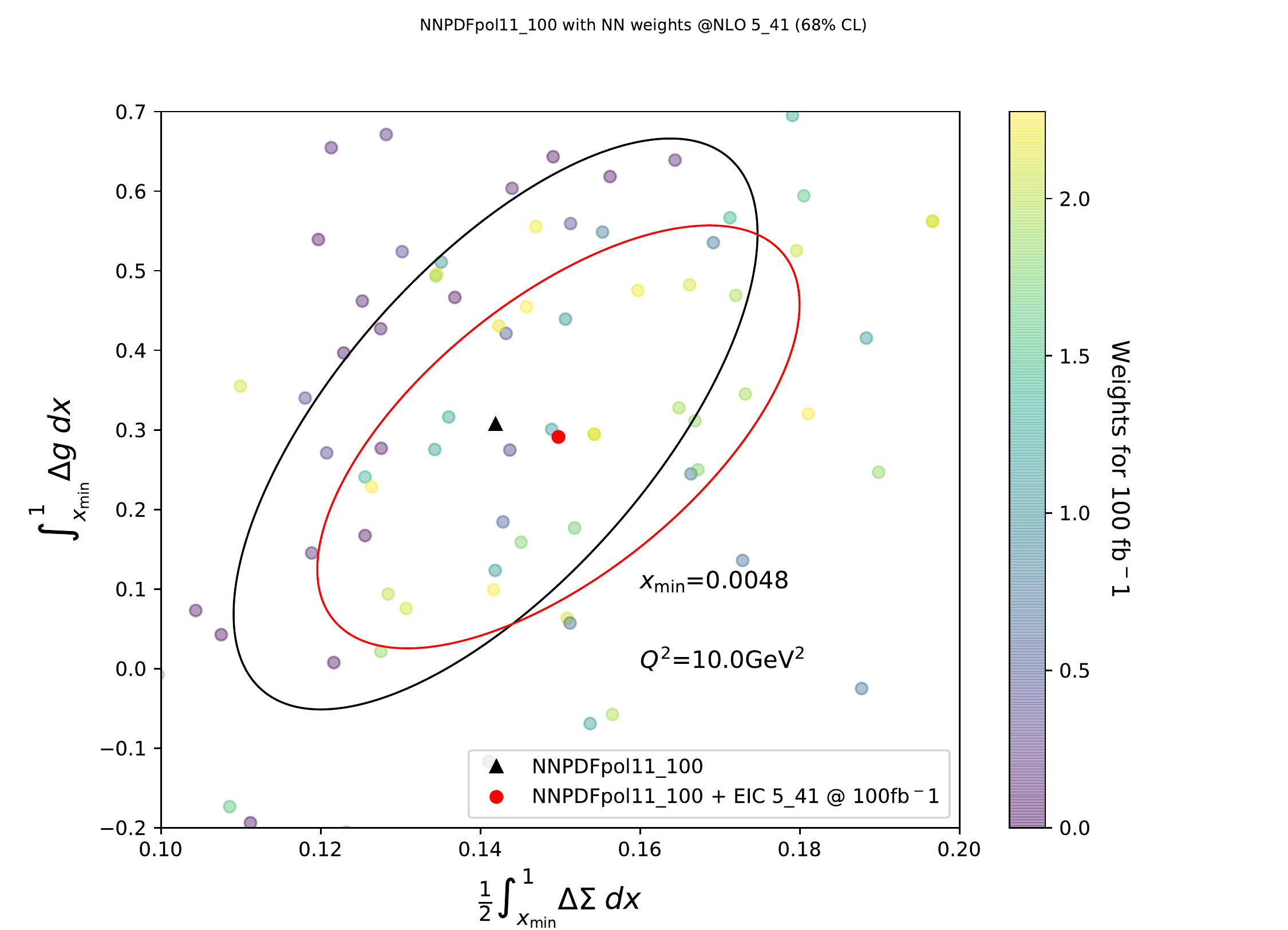}}
\\
\subfloat[For $\SI{5}{\GeV} \times \SI{100}{\GeV}$ energy configuration]{\includegraphics[clip, trim=0.5cm 0cm 2cm 1cm,width=0.45\textwidth]{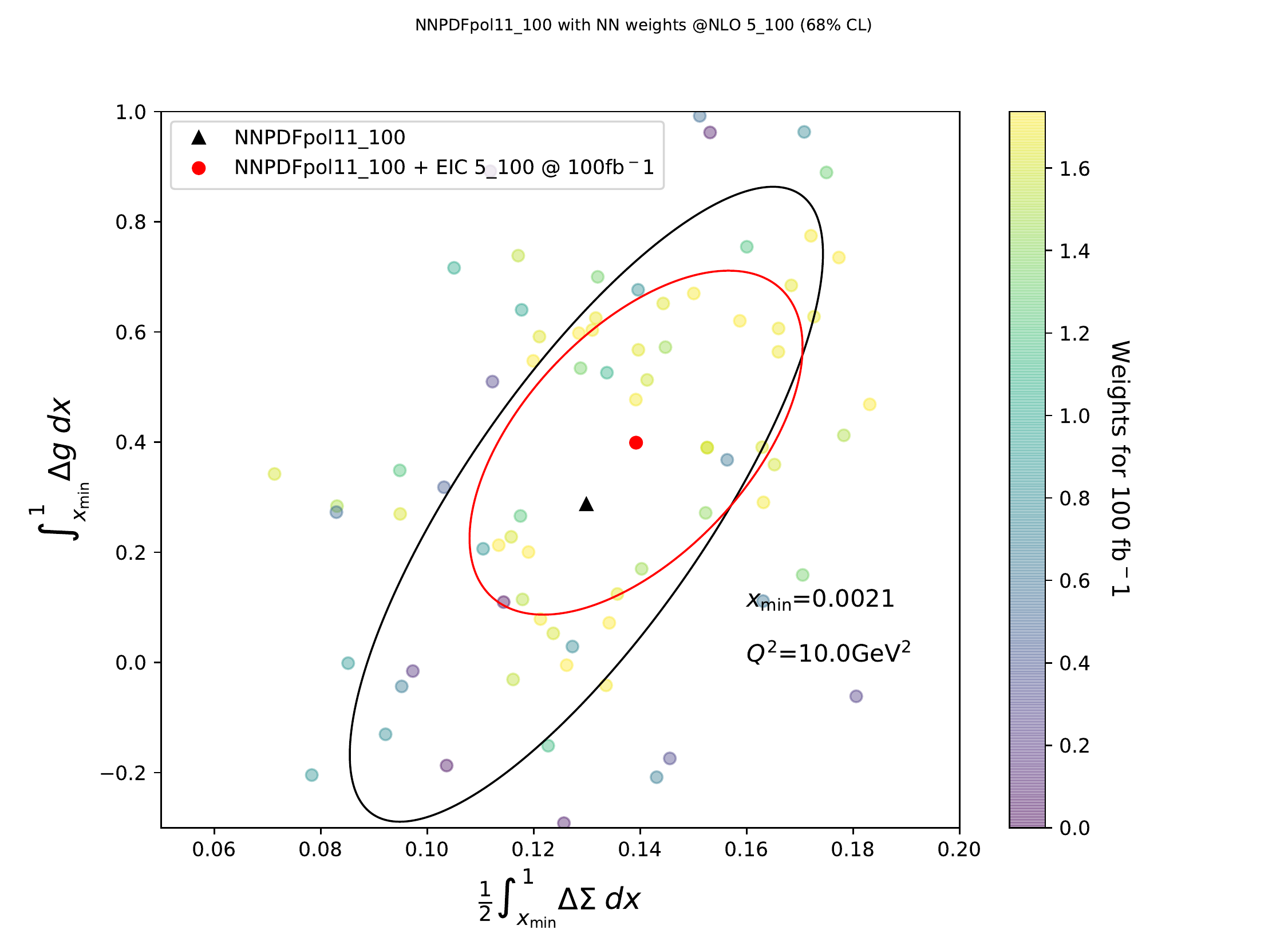}}
\\
\subfloat[For $\SI{18}{\GeV} \times \SI{275}{\GeV}$ energy configuration]{\includegraphics[clip, trim=0.5cm 0cm 2cm 1cm,width=0.45\textwidth]{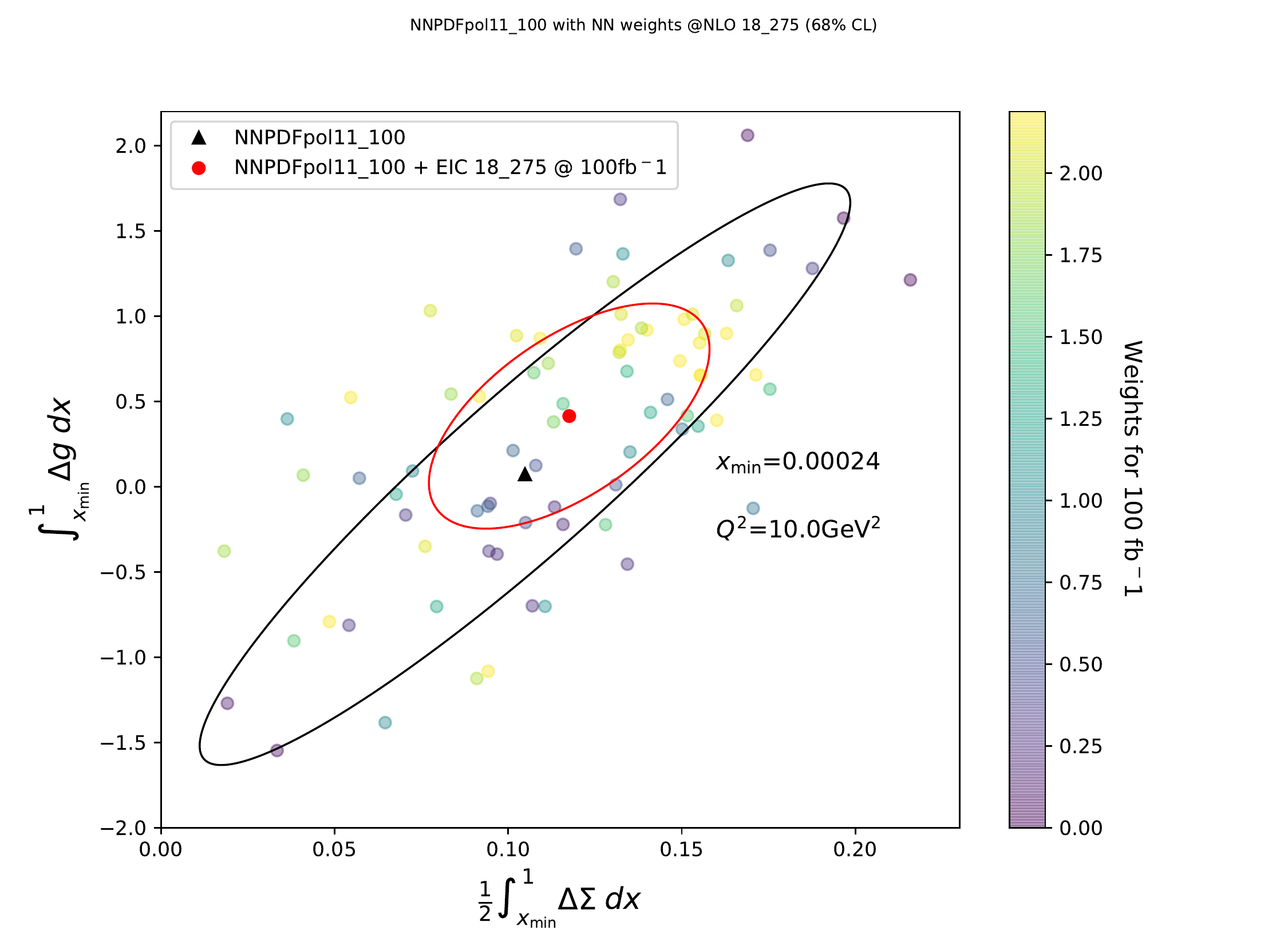}}
\caption{(Color online) Correlation plots at $Q^2=\SI{10}{\GeV^2}$ for the three different beam energy configurations. The original NNPDFpol1.1 68\% confidence level correlation ellipse is shown in black, whereas the red ellipses are the result of including different pseudo-data with integrated luminosity of \SI{100}{\per\femto\barn}. The weights associated with the replicas after including the pseudo-data are depicted using a color graded scale. For each collision configuration, the corresponding lower bound $x_\text{min}$ is shown in the plots. 
}  
\label{fig:correlation_NNPDF}
\end{figure}

\begin{figure}[htbp]
\centering
\subfloat[For $\SI{5}{\GeV} \times \SI{41}{\GeV}$ energy configuration]{\includegraphics[clip, trim=0.5cm 0cm 2cm 1cm,width=0.45\textwidth]{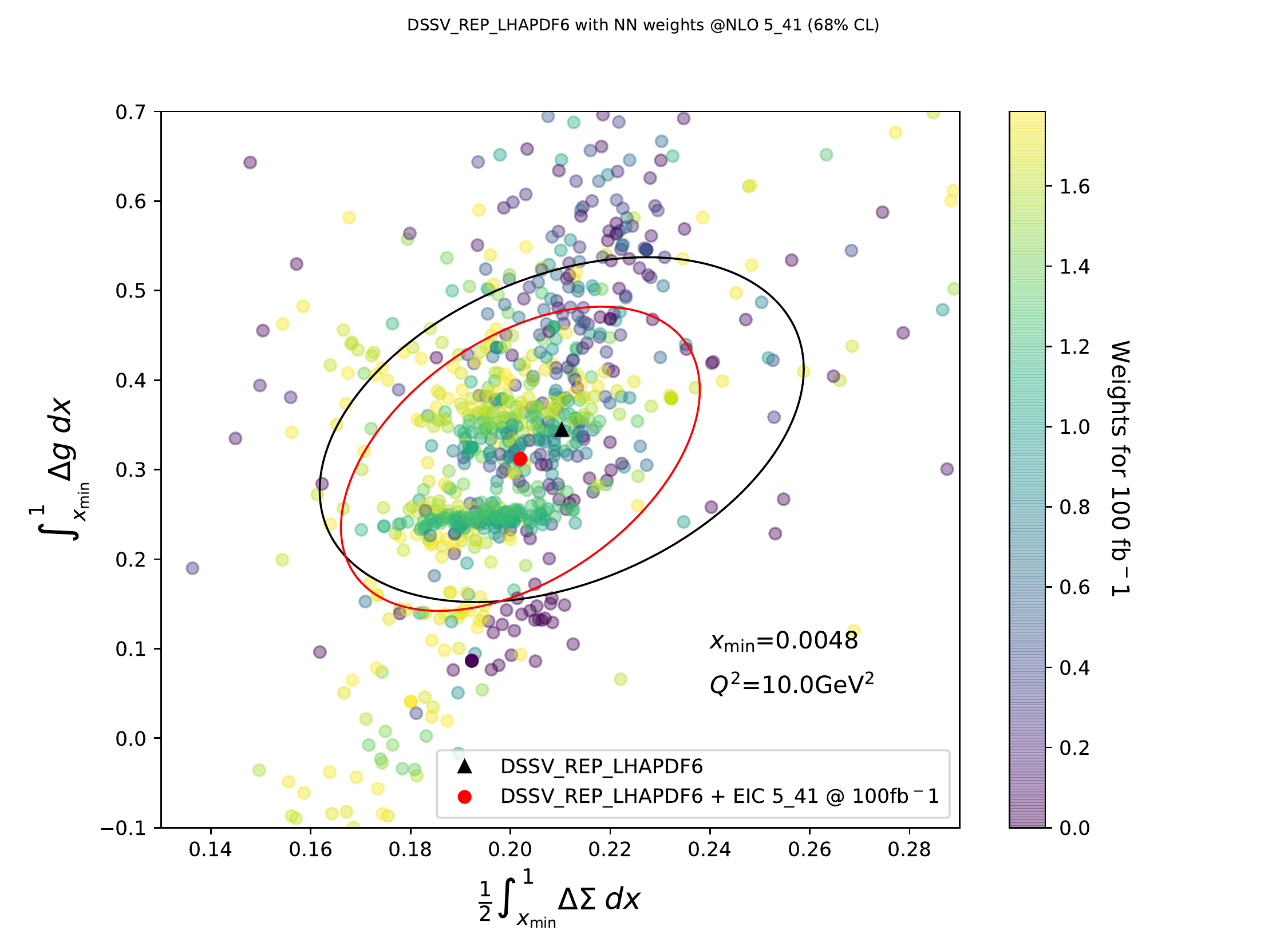}}
\\
\subfloat[For $\SI{5}{\GeV} \times \SI{100}{\GeV}$ energy configuration]{\includegraphics[clip, trim=0.5cm 0cm 2cm 1cm,width=0.45\textwidth]{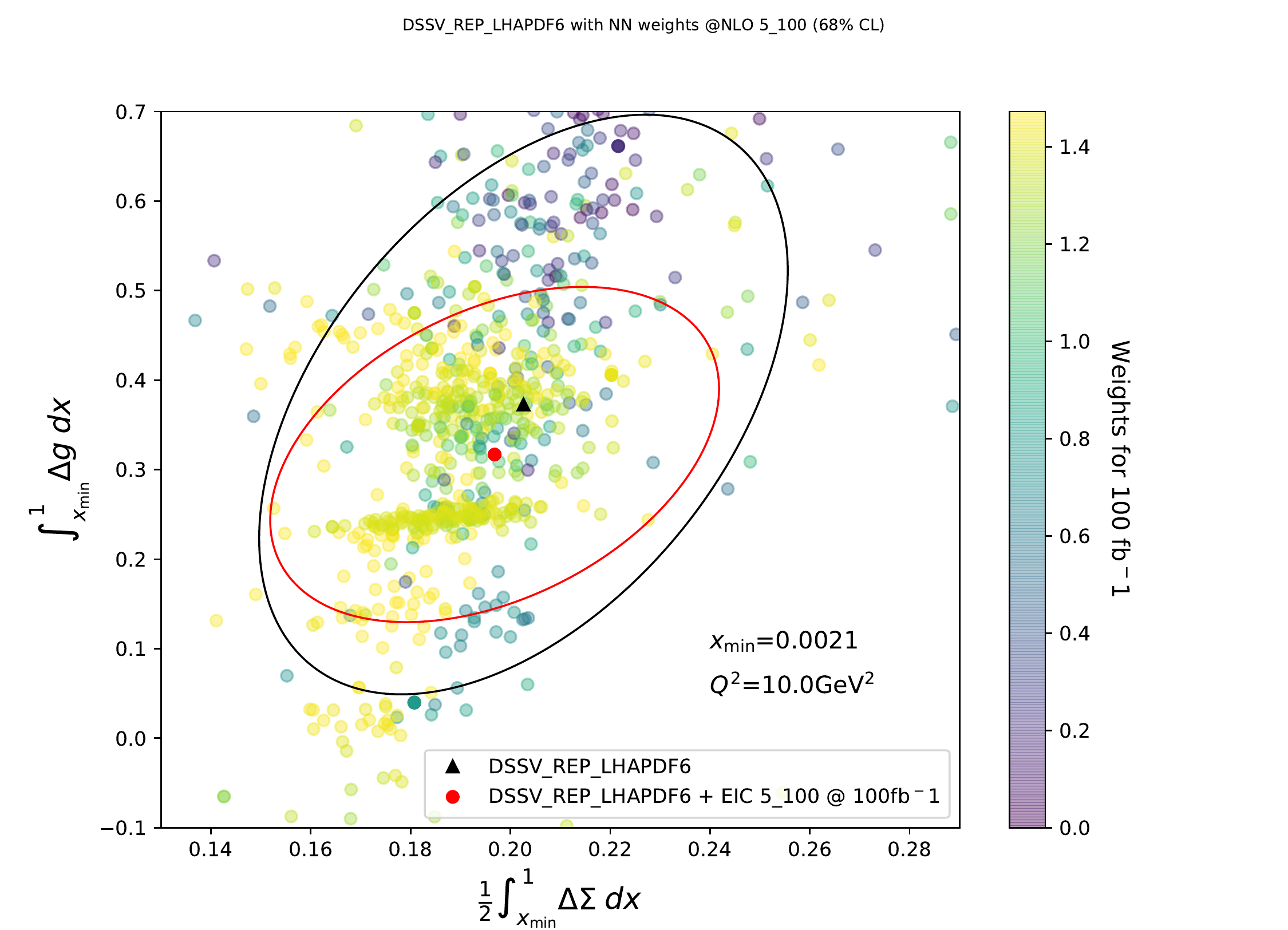}}
\\
\subfloat[For $\SI{18}{\GeV} \times \SI{275}{\GeV}$ energy configuration]{\includegraphics[clip, trim=0.5cm 0cm 2cm 1cm,width=0.45\textwidth]{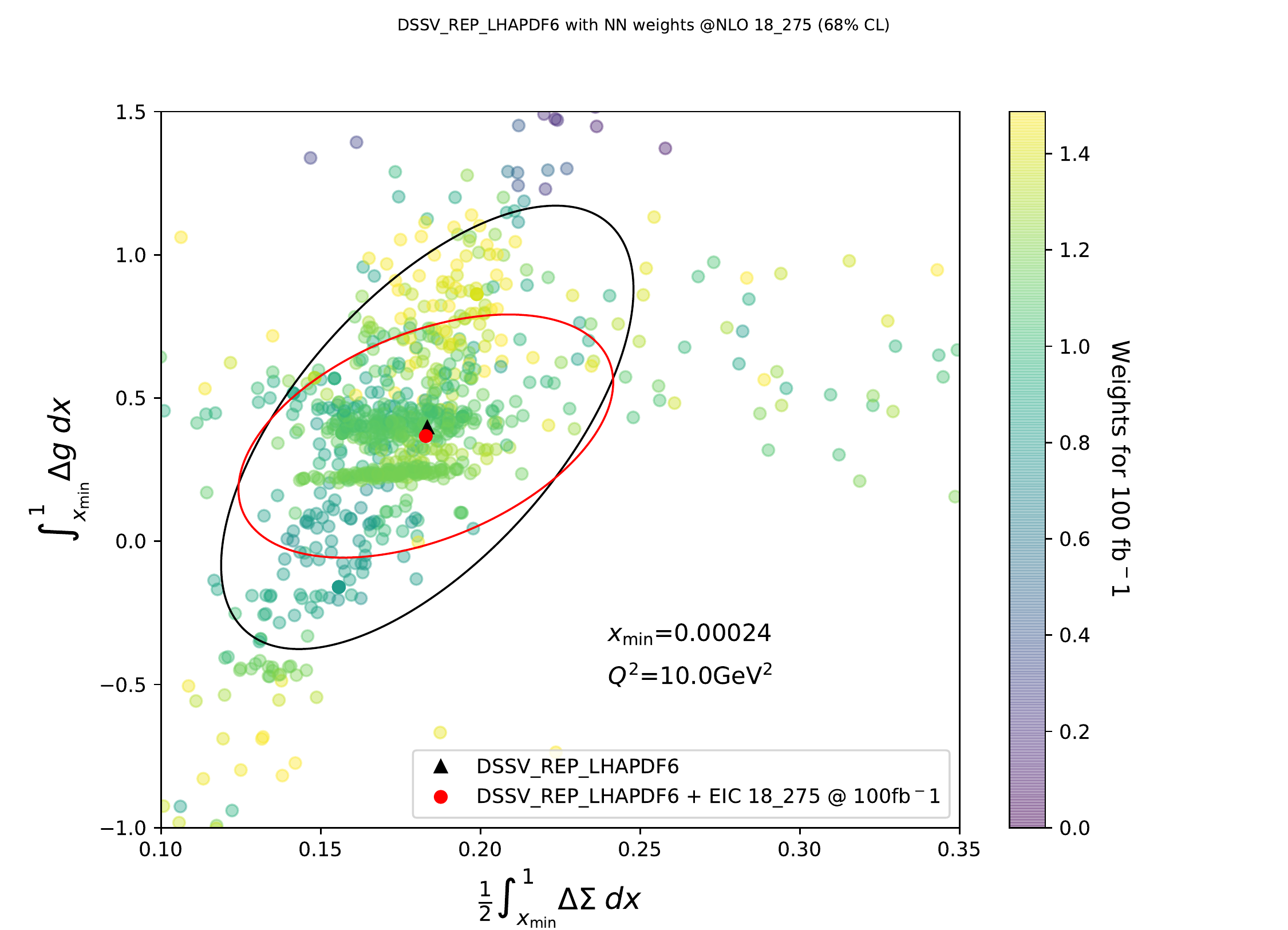}}
\caption{(Color online) Same as \cref{fig:correlation_NNPDF} for DSSV14 distributions. 
}  
\label{fig:correlation_DSSV}
\end{figure}

\section{Projections for the parton helicity distributions}
In global data fits, parton distribution functions are expressed in terms of some functional form depending on a number of free parameters whose values are constrained by the experimental data. A practical way of transferring this information to parton distribution functions themselves is to express them as a set of so-called ``replicas'' generated by means of Monte Carlo sampling of the parameter space. Central values and uncertainties of PDFs become then simple statistical mean values and standard deviations of the full replica set.
To assess the impact of the new measurement proposed in this paper on the quark and gluon helicity distributions without performing a full refit, we use the well-established reweighting method of~\cite{Ball:2010gb,*Ball:2011gg}. By exploiting Bayesian inference, the information contained in the new set of data can be incorporated directly into the probability distribution of the initial helicity PDF replicas. More specifically, this is achieved by assigning a weight to each replica measuring its consistency with the new data. The resulting new PDFs containing the information of the new data set preserve the statistical rigor of the original set as long as not too many replicas become suppressed by vanishing weights. 
A small number of surviving replicas would mean that the impact of the new data is too significant for the reweighting method to work and a full fit is necessary. 
In our study, two commonly used polarized PDF sets were used separately for the reweighting analysis: NNPDFpol1.1 \cite{Nocera:2014gqa} and DSSV14~\cite{deFlorian:2014yva,*deFlorian:2019zkl}. NNPDFpol1.1 is given in the form of 100 replicas while DSSV14 is given with 1000 replicas. 
Although replicas from both sets are generated using Monte Carlo sampling methods, the two sets profoundly differ in the way the shape of PDF replicas are parametrized. NNPDFpol1.1 uses functional forms provided through the use of neural networks, i.e. with a high number of free parameters. DSSV14 uses a more traditional analytical, although flexible, functional form with much fewer free parameters.

Without a real measurement, one doesn't know the central values of 
the data points. Therefore, the pseudo-data were generated by randomly
displacing the theoretical central values using the projected uncertainty.
Reweighting is performed using pseudo-data generated for the above-mentioned three different energy configurations and corresponding integrated luminosity of \SI{100}{\per\femto\barn}. 
Depending on the energy configuration, after reweighting the surviving replicas are around $\sim 70$ and $\sim 850$ for  NNPDFpol1.1 and DSSV14 respectively, which is a large enough number to justify the use of reweighting in this analysis.

In \cref{fig:impact_PDFs}, we show the impact of the EIC pseudo-data on the uncertainties of the singlet quark helicity distribution $\Delta\Sigma(x)$ and the gluon helicity distribution $\Delta g(x)$ at $Q^2=\SI{10}{\GeV^2}$ with \SI{100}{\per\femto\barn} of integrated luminosity. The gray band represents the original NNPDFpol1.1 or DSSV14 absolute uncertainty band before reweighting. Each colored band represents the effect of reweighting using pseudo-data generated with one of the three possible collision energy configurations and all of them combined. In the bottom area of the plots we also show the ratio between the uncertainties before and after reweighting. 
For the study with NNPDFpol1.1, one can clearly see the energy dependence of the pseudo-data
impact on the gluon helicity distribution: higher collision energy data offer more constraints on the pPDF uncertainty in the lower values of x, and vice-versa. In the quark sector, the two lower center-of-mass energy configurations have less constraining power. While for the study with DSSV14, the impact of
$\SI{18}{\GeV}\times\SI{275}{\GeV}$ and $\SI{5}{\GeV}\times\SI{100}{\GeV}$ pseudo-data sets is
similar in both quark and gluon sectors. Interestingly, the impact on $\Delta g$ is in the $x>0.1$ region for the $\SI{5}{\GeV}\times\SI{41}{\GeV}$ configuration, which is
a novelty of this measurement. 
The difference for the resulted impact of EIC pseudo-data by using NNPDFpol1.1 or DSSV14 is mainly due to two reasons.
The first one is due to the different data sets included in different global fits.
NNPDFpol1.1 fit contains the world data with $x$ extension to about $4\times 10^{-3}$ including DIS data, open-charm production data from
the COMPASS experiment at CERN and high-$p_{T}$ inclusive jet and $\pi^0$, as well as $W^{\pm}$ production
data from the STAR and PHENIX experiments at RHIC.
In addition to the DIS data, the DSSV14 also includes SIDIS data, inclusive jet and
identified hadron production measurements from polarized proton-proton collisions at RHIC.
The second reason is because of different parameterizations for the shape of quark and gluon helicity distributions.
One obvious effect of different parameterizations is the determined uncetainty band beyond the coverage of existing world data, for example in the very low-$x$ region, where the NNPDFpol1.1 shows a significant larger uncertainty compared to the DSSV14 case in both quark
and gluon sectors. 
That is why the high center-of-mass energy configurations show less impact in the relatively low-$x$ region for the DSSV14 study compared to the NNPDFpol1.1 case.
Note also that the $x$ value where the impact is large
extends beyond the Bjorken-$x$ reach of the data (see \cref{fig_ALL_projection}), this is
because of the shift between Bjorken-$x$ (determined by the virtual photon) and parton-$x$ ($x/z$ in~\cref{eq:F12,eq:g1}) in the 
PGF process.
The increase of the uncertainty band after reweighting for some values of $x$
in either quark or gluon sector is due to the fact that
the reweighting procedure is favoring replicas with an appropriate shape, 
which is mainly determined by the fixed parameterizations.
Similar reason also applies to the situation while looking at the impact by combining all the three pseudo-data sets, compared to the impact of individual
pseudo-data set.
Resolving such bias is outside the scope of this study and is better left to future fits with real EIC data.

When it comes to helicity PDFs, the first moments, i.e.\ their integrals over parton momentum fraction (see \cref{eq:Jaffe}) 
are quantities of great interest. They represent the net quark and gluon contribution to the proton spin~\cite{Jaffe:1989jz}. In every practical case, where the $x$ of the data is limited and the integration cannot be performed down to $x\rightarrow 0$, the truncated first moments 
are usually used to represent the contribution to the proton spin down to the $x_{\text{min}}$ fraction of proton momentum accessible by experimental data. 
In \cref{fig:xmin_NNPDF} and \cref{fig:xmin_DSSV} we present, for NNPDFpol1.1 and DSSV14 PDF sets respectively, the impact of different EIC pseudo-data sets on truncated first
moments for the quark and gluon helicity distributions as a function of the lower integration limit $x_{\text{min}}$. In the same way, we also show the missing contribution to the proton spin which is usually associated to the quark and gluon orbital angular momenta spin contributions~\cite{Jaffe:1989jz}. The bottom panel of each plot shows the
uncertainty improvement by including a particular set of pseudo-data.
Those plots are very instructive as they explicitly show which contribution to the integral is mostly affected by the data. From \cref{fig:xmin_NNPDF} we can observe that the two lower energy configurations are able to target specific regions of the gluon spin contribution to the proton spin. Choosing the $\SI{5}{\GeV}\times\SI{41}{\GeV}$ configuration, the precision of our knowledge of the contribution to the proton spin from gluons with momentum fraction down to the intermediate-$x$ region is increased up to a factor of 1.5. If we choose $\SI{5}{\GeV}\times\SI{100}{\GeV}$ configuration, this goes up to a factor of 2 around $10^{-3}\lesssim x_{\text{min}}\lesssim 10^{-2}$.
On the other end, the higher energy configuration is able to constrain the quark and gluon spin contributions to the proton spin in the $x_{\text{min}}$ region below 10$^{-3}$.
Using DSSV14 PDFs set (c.f. \cref{fig:xmin_DSSV}) leads to slightly different results. In particular, compared to the NNPDFpol1.1 case, we observe for the $\SI{5}{\GeV}\times\SI{41}{\GeV}$ configuration a larger impact in the high-$x_\text{min}$ and a general smaller impact of the $\SI{18}{\GeV}\times\SI{275}{\GeV}$ configuration for the quark sector.
This can be well understood by looking at \cref{fig:impact_PDFs}: if we compare the uncertainties of \cref{fig:impact_PDFs_NNPDF} versus \cref{fig:impact_PDFs_DSSV14}, the difference of impact of the lowest configuration energy in the quark sector between \cref{fig:xmin_NNPDF} and \cref{fig:xmin_DSSV} can be traced back to the difference in the original PDF uncertainties in the high-$x$ region. Similarly, the smaller impact in the small-$x_\text{min}$ region of the $\SI{18}{\GeV}\times\SI{275}{\GeV}$ data is related to the largely different uncertainties of NNPDFpol1.1 and DSSV14 in the low-$x$ region.
As gluon and quark sectors are typically correlated, we also present the first moments of gluon and quark helicity distributions in a two-dimensional plot. The plots in \cref{fig:correlation_NNPDF} and \cref{fig:correlation_DSSV} show the first moments of the gluon helicity distribution at $Q^2=\SI{10}{\GeV^2}$ as a function of the quark helicity distribution for the three collision energy configurations. For each plot $x_{\text{min}}$ is given by the lowest Bjorken-$x$ accessible to the specific energy configuration setting. The black ellipse represents the original 68\% C.L. uncertainty boundary before reweighting. The red ellipse represent the 68\% C.L. boundary when including pseudo-data with integrated luminosity of \SI{100}{\per\femto\barn}.
Each point in the plot is associated to a specific replica and their colors represent the magnitude of the weights associated after including pseudo-data. The dots in yellow represent the replicas dominating the distributions. They are clustered around the range of values that become relevant once EIC pseudo-data are inserted. The central position of this cluster is not physically relevant at this stage, as it is directly correlated with the distributions' central value shift which, as discussed above, takes meaning only once real experimental data are considered. 
We can notice that the cluster gets gradually squeezed as
the beam energy goes higher at an EIC. Moreover, for the study
with DSSV14 PDF sets, the ellipses after
including the EIC pseudo-data have different angles comparing to the original distribution, which means that the 
new observable offer independent ingredients, more specifically gluon-sensitive inputs, into the world data in the DSSV14 global
fit.


\section{summary}
We have proposed a new measurement on longitudinal double spin asymmetries in the $\vec{e} + \vec{p} \to e' + D^0 + X$
DIS process at an EIC to constrain the gluon helicity distribution $\Delta g$.
We would like to emphasize that the classic $g_1$ measurements at the EIC will play the dominant role to constrain the gluon helicity distribution and its contribution to the proton spin \cite{Accardi:2012qut}. Our proposal will provide complementary constraints on the gluon helicity distribution. As we show in the impact study, in some kinematics, e.g., moderate $x$ region, heavy flavor production will offer a unique opportunity. Especially, with a lower center-of-mass
energy machine, one can improve the precision of gluon helicity distribution in the $x>0.1$ region. 

Moreover, the theoretical calculation shows that the $A_{1}^{c}$ is sizable at the level of 10-20\% in the 
moderate $x$ region, which is an advantage experimentally. 
To achieve the measurement with good signal significance for the $D^0$ reconstruction, a state-of-the-art silicon pixel tracking detector system with
wide pseudo-rapidity coverage, excellent momentum and spatial resolutions as well as a low mass budget is essential.
It has been shown in our study and the work in~\cite{Arrington:2021yeb,Kelsey:2021gpk} that the proposed all-silicon tracker conceptual design is well suited for the measurement.

\section*{Acknowledgements}
The authors are grateful for the valuable discussions with Professor Nu Xu.
This work is supported in part by the Strategic Priority Research Program of Chinese Academy of Sciences, under grant number XDB34030301. D.\ A.\ and H.\ X.\ are supported by the Guangdong Major Project of Basic and Applied Basic Research No.~2020B0301030008, the Key Project of Science and Technology of Guangzhou (Grant No.~2019050001), the National Natural Science Foundation of China under Grant No.~12022512, No.~12035007. F.\ H.\ is  supported  by  the  European  Research  Council under  the  European  Unions  Horizon  2020  research  and innovation Programme (grant agreement number 740006).

\bibliographystyle{apsrev4-2}
\bibliography{references}
\end{document}